# Synthesis and Compression study of orthorhombic $Fe_7(C, Si)_3$: A possible constituent of the Earth's core

**Pinku Saha[1], Konstantin Glazyrin[2], and Goutam Dev Mukherjee[1*]**

[1]*National Centre for High Pressure Studies, Department of Physical Sciences, Indian Institute of Science Education and Research Kolkata, Mohanpur Campus, Mohanpur 741246, Nadia, West Bengal, India.*

[2]*Photon Science, Deutsches Elektronen Synchrotron, 22607 Hamburg, Germany*

*(Corresponding Author): goutamdev@iiserkol.ac.in

**Abstract**

The orthorhombic phase of Si-doped Fe carbide is synthesized at high pressures and temperatures using laser-heated diamond anvil cell (LHDAC), followed by its characterization using X-ray diffraction (XRD) measurements, Transmission Electron Microscopy (TEM), and Raman spectroscopy. High-pressure XRD measurements are carried out up to about 104 GPa at room temperature for determination of the equation of state (EOS) parameters of the synthesized sample. No evidence of structural transition is observed, though two anomalies are found in the compression behaviour of our sample at about 28 and 78 GPa, respectively. Pressure evolution of isothermal bulk modulus shows elastic stiffening around 28 GPa followed by softening around 78 GPa. These anomalies are possibly related to two different magnetic transitions driven by pressure-induced anisotropic strain in the unit cell. Extrapolation of the density profile of our study to the inner core conditions agrees very well with PREM data with an uncertainty of about 3-4%. We have estimate bulk modulus value seems to be 8-9% less than that of PREM data in the shown pressure range and is best matched in comparison to other reported values for the non-magnetic phase.

**Keywords:** Laser heated diamond anvil cell, orthorhombic Si-doped $Fe_7C_3$, elastic anomalies, Earth's inner core density, crystal structure

# 1 Introduction

For several decades, the composition of Earth's core has been under extensive debate in scientific communities. Amongst several elements, Iron (Fe) and its alloys with Nickel (Ni) have been predicted to be the main components of the Earth's core from X-ray diffraction experiments at extreme pressures. But densities (Mao et al., 1990; Dubrovinsky et al., 2000) of these materials are found to be significantly higher than that of Earth's core estimated from seismic observations (Dziewonski & Anderson 1981; Stevenson 1981). Given the above results, it has been suggested that a few percentages of light elements may be present along with Fe or Fe − Ni alloy (Birch 1952; Wood 1993; Poirier 1994; Li et al., 2002; Li & Fei 2003). Initially, Carbon (C) was considered to be the leading light element along with Fe in the form $Fe_3C$ at Earth's core due to its high solar abundance, chemical affinity to iron at low pressures, and overall ability to lower the density of pure Fe or Fe − Ni alloy (Wood 1993). Sata et al. (2010) show that C is the most dominant constituent along with Fe among other light elements such as Si, O, S due to its minimal density deficit concerning the PREM data at inner core conditions (Dziewonski & Anderson 1981). However, theoretical studies predicted a non-magnetic $Fe_3C$ with a larger bulk modulus compared to the PREM data (Vočadlo et al., 2002).

High pressure and high-temperature experiments using both multi-anvil (MA) cell and LHDAC on Fe − C system resulted in the formation of a new iron-carbide phase with chemical formula $Fe_7C_3$ at about 1500°C and 10 GPa, which is predicted to be a potential candidate for the solid inner core (Nakajima et al., 2009; Nakajima et al., 2011; Mookherjee et al., 2011; Chen et al., 2012; Chen et al., 2014; Prescher et al., 2015; Liu et al., 2016). Several research groups indexed the synthesized phase to the hexagonal structure (Nakajima et al., 2011; Chen et al., 2012). Electronic structure calculations by Mookherjee et al. (2011) on hexagonal $Fe_7C_3$ at

Earth's core conditions proposed around 7% less density as compared to PREM data. Experimental work by Chen et al. (2012) proposed the density of hexagonal $Fe_7C_3$ be about 5-10% lower concerning the PREM data at inner core pressure and temperature range 5000-7000 K. Recently, Prescher et al. (2015) synthesized an orthorhombic phase of $Fe_7C_3$ using MA apparatus. A theoretical simulation study by Das et al. (2017) comparing the hexagonal, orthorhombic pure phases of $Fe_7C_3$ and silicon (Si) doped $Fe_7C_3$ showed that values of density and Poisson's ratio of orthorhombic ($o$)-$Fe_7(C, Si)_3$ at the inner core to be very close to the PREM data. The above study proposed $o$-$Fe_7(C, Si)_3$ with 3.2 wt % of Si at C cite to be one of the most important components of the Earth's inner core.

Depending on the current trends of the possible candidates for the core material, we have successfully synthesized the orthorhombic phase of Si-doped $Fe_7C_3$ at high pressures and high temperatures using the LHDAC facility present in our laboratory. We have characterized the above material by TEM, Raman, and XRD measurements. We have also carried out the equation of state measurements on Si-doped $Fe_7C_3$ at room temperature. We have estimated the bulk modulus and the density at inner core conditions by extrapolating our results, which is found to be closest to the PREM data within error limits.

## 2 Material and Experimental Methods

We have prepared a homogeneous mother stoichiometric mixture of 2 gm taking the appropriate amount of powders of Fe (CAS No: 7439-89-6, purity: >99.5% (RT), grain size: 5-9 micron), nano-diamond (CAS No: 7782-40-3, purity: >95% trace metal basis, grain size : 10 nm) and Si (CAS No: 7440-21-3, purity: >99% trace metal basis, grain size: 2-5 micron) having atomic ratio Fe : C : Si = 70:25:5, which is ground over a long period of time in alcohol. This provides us a homogeneous mixture of the sample. The thin pellets of the above mixture of the approximate size of about 70 - 100 μm, sandwiched between pellets of dried NaCl (~15 – 20 μm thickness) are loaded in LHDAC for *in-house* experiments. The LHDAC consists of a plate-type DAC (Boehler-Almax design) of culet diameter 300 μm and a stainless steel gasket with a hole of diameter ~ 110 μm pre-indented to a thickness of 50 μm. A few ruby chips of the approximate size of 5 μm are placed at the gasket edge to determine the pressure in the LHDAC using the ruby fluorescence method (Mao et al., 1986). Laser heating of the sample is carried out using IPG photonics CW laser of wavelength 1.07 μm (maximum power of 100 Watt). The temperature of the sample is measured using the spectra-radiometry technique (Boehler et al., 1990; Mukherjee & Boehler 2007) by fitting Planck's radiation function (Planck 1901) to the collected flatfield corrected spectra as described in the earlier studies (Saha & Mukherjee 2020; Saha et al. 2020). Three different heating runs are carried out on the sample at pressures of 8, 15, and 28 GPa at a hotspot temperature of about 2000 K. Heating is carried out for 2-5 minutes in each run. XRD measurements of the retrieved samples are carried out in the P02.2 beamline of the Petra-III synchrotron radiation facility in Hamburg, Germany. Raman measurements of the unheated and heated samples (after washing the retrieved sample using distilled water) are carried out in backscattering geometry using the micro-Raman spectrometer (Monovista CRS + from S&I) with a grating of 1500 grooves mm$^{-1}$, 20X long working distance objective (infinitely corrected), and the excitation wavelength of 532 nm with a spectral resolution of 1 cm$^{-1}$.

*In situ* high-pressure high-temperature X-Ray diffraction measurements are also carried out in the P02.2 beamline of Petra-III synchrotron source using the LHDAC facility consisting of membrane driven symmetric DAC of culet flat ranging 150-300 μm. In this case, dried MgO is used as a pressure transmitting medium as well as a pressure marker (Jacobsen et al., 2008). A total number of four different high-pressure high-temperature XRD measurements are performed. In all LHDAC heating experiments, the laser spot is moved around in pre-determined grids for heating of the complete sample volume. We have carried out the heating run for 2-5 minutes at each pressure and temperature and continuously checked the evolution of patterns during heating. Synchrotron powder diffraction measurements are carried out using monochromatic X-ray radiation of wavelength 0.29 Å collimated to an area of 1.2×2.3 $\mu m^2$. Diffracted X-rays are detected using a Perkin-Emler 1681 detector aligned normal to the beam. The sample to detector distance is calibrated using the $CeO_2$ standard. Collected two-dimensional diffraction images are integrated to 2θ versus intensity profile using Dioptas software (Prescher & Prakapenka 2015) and then analyzed using CRYSFIRE (Shirley 2002) and Rietveld refinement program GSAS (Toby 2001). Details of high-pressure, high-temperature sample synthesis conditions using LHDAC, and XRD measurements are tabulated in Table-1.

After high pressure and high-temperature treatment in the P02.2 beamline, the sample is retrieved and prepared for TEM measurements as described below. The retrieved sample is enclosed within the MgO pressure transmitting medium. MgO is removed very carefully under a microscope. Then the retrieved sample is polished to make it very thin, suitable for TEM measurements. Chemical mappings of the retrieved samples are carried out using the TEM operating at 200 kV. We have recorded bright-field electron image and electron diffraction at selected areas to determine the crystal structure using the same instrument.

**3 Results and Discussion**

   A. **Sample at ambient Condition:**

Scanning electron microscope (SEM) imaging and elemental mapping of the starting mixture are shown in Supplementary Figure S1. Elemental mapping and spectra measurements are carried out from 4×3 μm$^2$ rectangular cross-section shown by the magenta rectangle in the electron image, revealing the presence of the Fe, C, and Si in the starting mixture. Raman spectra of unheated and recovered samples after heating *in-house* at three different runs are shown in Supplementary Figure S2. Raman spectrum of the unheated sample reveals three modes centered at 520, 1330, and 1580 cm$^{-1}$ corresponding to Si, D-band and G-band of nano-diamond, respectively. But Raman spectra of the heated sample do not show any prominent mode of the parent constituents. Disappearances of the Raman bands after heating exclude the presence of un-reacted elemental C and Si in the heated samples. Supplementary Figure S3 shows the XRD patterns of the quenched samples retrieved from the *in-house* experiments. 2-D (two-dimensional) diffraction images of the synthesized sample in different runs show the spotty nature of the image indicating preferred orientated crystal growth. However the presence of ring patterns in the image prompted us to carry out the Le-bail profile fitting of the patterns. The XRD patterns are best indexed to an orthorhombic structure. Synthesis conditions and the lattice parameters are tabulated in Table-1. Volumes determined from different synthesized samples are observed to differ by a maximum of 0.25%, and the average volume is found to be 819 (5) Å$^3$. *In situ* study at high-temperature and high-pressure on iron also has revealed an orthorhombic structure (Andrault et al., 1997). Le-bail profile fittings using the orthorhombic structure show very good fits for samples synthesized at 15 and 28 GPa as evident from Figure S3 (a & b). However, the sample synthesized at a low pressure of 8 GPa shows relatively poor fitting to the orthorhombic structure, showing the possibility of stabilization of the orthorhombic phase at relatively higher pressures. In comparison to the pure orthorhombic phase of Fe$_7$C$_3$ at room temperature and pressure as reported in the literature (Prescher et al., 2015; Liu et al., 2016), we have observed 4%, 6%, and 10% increase in the values of *a*, *c*, and *V*, respectively; while we obtain a similar value of *b*. Possibly, the larger diameter of Si (2.10 Å) to that of C (1.70 Å) and low electronegativity of Si (1.9) to that of C (2.55) resulted in the expansion in the lattice parameters and hence in volume. All the above experimental results reveal that the samples synthesized *in-house* belong to orthorhombic crystal structure with space group *Pbca*, where elemental Fe, C, and Si are absent.

We have carried out four different HP-HT XRD measurements on the pressed mixtures at the P02.2 beamline and will discuss the results in detail later. First, let us discuss the elemental mapping of the synthesized sample. For mapping of the chemical composition of the quenched samples in the above runs, we could prepare the retrieved samples of two runs, Run#3 and Run#4 (Table-1) for TEM. The mappings reveal the presence of Fe, C, and Si in the synthesized compound at an atomic ratio very close to the starting material and are shown in Supplementary Figure S4. Bright field electron image and selected area diffraction pattern (SADP) of the retrieved sample from Run#4 are shown in Supplementary Figure S5. Analysis of SADP (Lin 2014) reveals the orthorhombic phase of silicon doped $Fe_7C_3$ in space group *Pbca*.

### B. Sample at high pressures:

Pressure evolution of XRD patterns in the first three runs are shown in the Supplementary Figures S6 to S8. In each run, we have also shown the Le-bail fitting of two patterns at different pressures in the same figure, respectively. The synthesis conditions and the fitted parameters are presented in Table-1. We have observed different intensities of the Bragg planes in the patterns of different runs, which may be due to the strong preferential orientations of the crystallites formed at different pressure and temperature conditions. But all the patterns could be fitted very well to the orthorhombic crystal structure in space group *Pbca*. XRD pattern of the hexagonal phase of $Fe_7C_3$ (Nakajima et al., 2011) is similar to the pattern observed in Run#3 and #4 in this present study. In the present case, the hexagonal structure resulted in a poorer fit in comparison to the orthorhombic structure. In Table-2, the experimentally observed values of $2\theta$ are compared with those values obtained from the best fit to hexagonal and orthorhombic structure at three different pressures. The deviations in the $2\theta$ values calculated using hexagonal phase lattice parameters are found to be larger compared to those corresponding to the orthorhombic phase. Comparison of the intensity profile among experimental, computed hexagonal and computed orthorhombic structure at three different pressure points are shown in Supplementary Figure S9. It may be noted that experimental patterns show a better match to the computed orthorhombic structure. Therefore XRD analyses in combination with the SADP analysis indicate that our synthesized samples are in the orthorhombic phase (space group *Pbca*). To check the structural homogeneity we have carried out in situ XRD measurements just after heating at different positions of the synthesized samples using a preselected grid pattern (Figure S11(d)). In Supplementary Figure S11, we have shown XRD patterns of the sample obtained at selected runs. The positions of the Bragg peaks in all the patterns reveal that the synthesized sample is structurally homogeneous. Certain mismatch in the peak intensities are again due to the preferred orientation of the crystals at different positions and are captured due to the small size of the XRD beam.

In Figure 1(a and b), we have shown the XRD patterns of the synthesized sample in Run#4 at 49, and 90 GPa along with Le-bail fitting. The large intensities of MgO Bragg peaks compared to our sample are mainly due to the fact that we have loaded thin pressed pellets of the sample inside the gasket hole mostly filled by MgO pressure transmitting medium. This reduces pressure inhomogeneity across the sample volume inside the LHDAC. The evolution of the patterns is shown in Figure 1(c). Henceforth, we label our synthesized orthorhombic Si-doped $Fe_7C_3$ as $o$-$Fe_7(C, Si)_3$. *In-situ* X-Ray diffraction measurements at higher pressures and temperatures in P02.2 beamline confirmed the presence of the same phase of $o$-$Fe_7(C, Si)_3$. Our high pressure XRD measurements in four different Runs on $o$-$Fe_7(C, Si)_3$ at room temperature up to about 104 GPa do not show any structural phase transition. Pressure evolutions of volume in four different runs at room temperature are shown in Supplementary Figure S10. We find three different regions: from ambient to 28 GPa – range(i), 28 to 78 GPa – range(ii), and above 78 GPa – range(iii), which will be discussed in the next paragraph. Pressure versus volume data are fitted to the 3$^{rd}$ order BM EOS and the values of the bulk modulus and their pressure derivatives in all Runs are presented in the figure. Ambient volume is taken to be 819 Å$^3$ determined from the analysis of the XRD patterns of the *in-house* retrieved samples for range (i) and (iii). In the range (ii) of Runs #2, #3, and #4 the above ambient volume fails to fit observed PV data. Therefore the ambient volume is estimated by the polynomial fit to the data. All the results are well in agreement within the error bars. We consider Run #1 and Run #4 for the rest of this manuscript.

In Figure 2(a) we have compared the pressure evolution of the volume of the $o$-$Fe_7(C, Si)_3$ at room temperature of Run #1 and #4 in our study with those reported in the literature for pure $o$-$Fe_7C_3$ (Prescher et al., 2015) and Si-doped $o$-$Fe_7C_3$ (Das et al., 2017). We have found 1-3% volume increment at high pressure (28-104 GPa) regime in our study compared to the above studies. The errors in our data are calculated using the error obtained in pressure from the MgO pressure marker. Volume error is obtained from the indexing of XRD patterns. To understand the effect of strain on the compression behavior of $o$-Fe$_7$(C, Si)$_3$, we have plotted the reduced pressure $\left(H = \frac{P}{3f_E(1+2f_E)^{5/2}}\right)$ versus the finite Eulerian strain $\left(f_E = \frac{1}{2}\left[\left(\frac{V_0}{V}\right)^{2/3} - 1\right]\right)$ (Polian et al., 2011) in Figure 2(b). The behaviour is expected to be linear following the equation (Murnaghan 1937; Birch 1947):

$$H = K_0 + \frac{3}{2}K_0(K' - 4)f_E \tag{1}$$

where, $V_0$ is the volume at 1 bar and 300 K, $V$ is the volume at pressure P, $K_0$ is the bulk modulus, and $K'$ is the first derivative of the bulk modulus. However, the plot shows two distinct discontinuities at around 28 GPa and 78 GPa, respectively. In the pressure range 28-78 GPa the Eulerian strain shows almost 70% increment, indicating a large pressure induced strain in the lattice. The discontinuities in the $H$ versus $f_E$ plot have been previously observed in Fe$_3$C and $h$-Fe$_7$C$_3$ system at the magnetic transitions and with no structural transition (Nakajima et al., 2011; Chen et al., 2012; Prescher et al., 2012). We have carried out separate 3$^{rd}$-order B-M EOS fitting in the regions (i) 0-28 GPa, and (ii) 28-78 GPa, and are shown in Figure 2(a). The fitted values of $V_0$, $K_0$ (bulk modulus) and $K'$ (first derivative of the bulk modulus) in the pressure range (i) are 819(5) Å$^3$, 70(8) GPa and 4.9(7), respectively. We have obtained similar values of the $K_0$ and $K'$ while fitting of $H$ vs. $f_E$ plot in the range (i) taking $V_0 = 819(5)$ Å$^3$. The value of $K'$ is consistent with ε-Fe and $o$-Fe$_7$C$_3$ (Chen et al., 2012; Prescher et al., 2015). The sample seems to be highly compressible up to 28 GPa. In the range-(ii), values of $V_0$, $K_0$ and $K'$ are found to be 714(6) Å$^3$, 298(17) GPa and 6.2(6), respectively. $K_0$ and $K'$ are consistent with the hexagonal phases of Fe$_7$C$_3$ in their paramagnetic phase (Nakajima et al., 2011; Chen et al., 2012; Liu et al., 2016). In this pressure range, the sample is less compressible than the range (i). Above 78 GPa, the volume decreases faster with pressure indicating an increase in the compressibility. For being a very narrow range of pressure and unavailability of $V_0$ in the range (iii) 78-104 GPa, we fit our $H$ vs. $f_E$ plot for data to estimate $K_0$ and $K^{'}$. The fitted value of $K_0$ is 152(4) GPa and $K^{'}$ is 3.8(4). We use $3^{rd}$-order Birch-Murnaghan EOS and find the best fit at $K_0$ is 150(7) GPa and $K^{'}$ is 3.9(2) for $V_0$ = 819(5) Å$^3$. The value of $K_0$ seems to be low in comparison to that of the paramagnetic phase of $h$- $Fe_7C_3$ (Nakajima et al., 2011; Chen et al., 2012; Liu et al., 2016). However, the value of $K^{'}$ is consistent with the non-magnetic phases of iron carbide (Chen et al., 2012; Mookherjee et al., 2011). The effect of ambient volume ($V_0$), $K_0$, and $K^{'}$ on the pressure-dependent volume will be discussed in the next paragraph. In Figure 2(c) we have shown the pressure evolution of individual lattice parameters in this study along with literature values for pure and Si-doped $o$-$Fe_7C_3$ (Prescher et al., 2015; Das et al., 2017). Interestingly $a$ and $c$ axes show pressure behaviour similar to volume, but the $b$-axis shows a single anomaly at 28 GPa. This behaviour indicates an anisotropic compression of the unit cell. The axial compression behaviour $\left(\beta_0 = \frac{(\delta l/\delta P)_{P=0}}{l_0}\right)$, where $l_0$ is the lattice parameters at 1 bar and 300 K for each axis is estimated by fitting pressure dependence of each axis to a $3^{rd}$-order B-M EOS (Angel et al., 2000). A single EOS function could not be used to fit the data in the complete pressure range: (1) three distinct EOS parameters for $a$ and $c$ - axes and (2) two distinct EOS parameters for $b$- axis. In the pressure range 0 - 28 GPa, $b$- axis seems to be least compressible. In the pressure range 28 - 78 GPa, the compressibility of $c$- axis reduces drastically compared to the other two axes. At high pressures above 78 GPa, the compressibility along all the axes almost behave similarly. Such different linear compressibility has not been reported in the literature for other iron-carbide phases (Nakajima et al., 2011; Chen et al., 2012; Prescher et al., 2015; Liu at al., 2016). The values of linear compressibility (β) along three lattice parameters at three different regions are presented in Supplementary Table-S1. High-pressure Mössbauer spectroscopic study of $Fe_3C$ shows a ferro-to-paramagnetic transition around 8-10 GPa and a spin transition at about 22 GPa (Prescher et al., 2012). Other high pressure experiments on hexagonal $Fe_7C_3$ have shown two anomalies in the compression curve at around 7-18 GPa and about 53 GPa corresponding to the ferromagnetic to paramagnetic and paramagnetic to non-magnetic phase transition, respectively (Nakajima et al., 2011, Chen et al., 2012). In $o$-$Fe_7C_3$, Prescher et al. (2015) have observed a ferromagnetic to paramagnetic phase transition at around 16 GPa and a paramagnetic to non-magnetic phase transition at 70 GPa, which is reflected in the mean C to Fe atom distance with pressure. In all possibility, these magnetic transitions are reflected in our volume as well as axial compressibility behavior.

## 4. Implications

Earth's core is at a much higher pressure than the maximum pressure obtained in the present study. To determine the density of materials at core pressures one needs to extrapolate our data points and apply the effect of temperature accordingly. Due to the unavailability of the ambient volume of the range (iii), we take our $V_0$ value as ambient volume and fit range (iii) to $3^{rd}$ order B-M EOS as shown in Figure 2(a) and Figure 3(a). Considering no structural transition in the whole range this is a good approximation to start with. We have also tried to fit the data points taking the ambient volume from the literature of pure and Si-doped $o$-$Fe_7C_3$ (Prescher et al., 2015; Das et al., 2017). Best fit for $V_0$ of pure and Si-doped $o$-$Fe_7C_3$ produce $K_0 = 613(26)$ GPa, $K^{'} = 0.1(2)$, and $K_0 = 328(16)$ GPa, $K^{'} = 2.1(2)$, respectively. But both of these fitted curves show discontinuity at high pressure, which restricts the use of these parameters for extrapolation. We have found systematic behaviour of fitted curves at high pressure with values $K_0 = 282(15)$

GPa, $K^{'} = 3.1(3)$ for pure, and $K_0 = 465(18)$ GPa, $K^{'} = 2.4(2)$ for Si-doped $o$-Fe$_7$C$_3$, those are shown in Figure 3(a). It can be noted that these curves show a very poor fit to our data points at both ends. So we advance with our $V_0$ value as ambient volume. We have fitted data points with varying $K_0$ from 130 to 170 GPa, which yields different values of $K^{'}$. The observed correlation is shown in Figure 3(b), a monotonic decrease of $K^{'}$ value is observed with increasing $K_0$. A few fitting curves obtained by varying $K_0$, and $K^{'}$ are shown in Figure 3(c) and these are extrapolated to Earth's core pressure 360 GPa. Inset is the zoomed-in view of the region that is marked by a red rectangle. Inconsistency of the fitted curves can be noted from the inset for $K_0$ values above 160 GPa, and below 135 GPa (corresponding $K^{'}$ limit 3.56, and 4.48, respectively). The volumes above 104 GPa, the highest pressure in our study are obtained by extrapolating our fitted EOS parameters of range (iii). The error in the extrapolated volumes are assigned from the deviation of the fitted curves related to $K_0 = 160$ GPa, and 135 GPa from that related to $K_0 = 150$ GPa. Error in the determination of the volume increases with pressure and the maximum error is about ±2.5% at 360 GPa.

We have estimated the density of $o$−Fe$_7$(C, Si)$_3$ from 104 GPa to inner core pressure at 300 K taking into account the EOS parameters of the range-(iii) in our study and compared with that of pure Fe, PREM data, and other phases of Fe$_7$C$_3$ in Figure 4 (Chen et al., 2012; Dubrovinsky et al., 2000; Dziewonski & Anderson, 1981; Liu et al., 2016; Nakajima et al., 2011; Prescher et al., 2015). We have assigned the errors in the density (Figure 4) from the errors in the measurement of the mass of the mother mixture at ambient pressure and errors in the volume as explained above. The overall error in the density measurement is found to be ± 3.2% at the core pressure. We have found around 4% higher density of our $o$−Fe$_7$(C, Si)$_3$ at 300 K in comparison to PREM data. The inner core temperature ranges from 5000 to 7000 K as estimated by Boehler

(1996). We have estimated the temperature effect on our density value from the thermal pressure following the equation $P(V, T) = P(V, T_0) + P_{th}(T)$; where $P(V, T_0)$ is the B-M EOS and $P_{th}(T)$ ($\alpha K_T \Delta T$, where α is thermal expansion coefficient and $K_T$ is the isothermal bulk modulus and $\Delta T = T - T_0$) is the thermal pressure (Anderson, 1984). Thermal pressure $P_{th}(T)$ is the pressure that would be created by increasing temperature from $T_0$ to T at constant volume. Since we have used MgO as pressure standard and observed the pressure before heating, during heating and after heating around 2000 K in a few runs, we can assess the effect of temperature from thermal pressure on density. We have plotted temperature versus thermal pressure and shown in Figure S12 in the Supplementary information. We have found 10 GPa of thermal pressure at a temperature 2000 K from MgO EOS, which in turn provides, $\alpha K_T$ = 0.0058 (5) GPa$^{-1}$. Extrapolation of the density profile of the range (iii) in our study to the inner core pressures at 5000 K is found to 1.5% higher than PREM data (Dziewonski & Anderson, 1981) and at 7000 K it matches very well. Among all the reported densities at the Earth's core pressures and temperatures, extrapolation of the density profile of the range (iii) at 7000 K to the inner core pressures agrees very well with the PREM data (Chen et al., 2012; Dubrovinsky et al., 2000; Dziewonski & Anderson, 1981; Liu et al., 2016; Nakajima et al., 2011; Prescher et al., 2015).

The elastic hardening of $o$-Fe$_7$(C, Si)$_3$ is observed at 28 GPa as indicated by a jump of bulk modulus ($K_0$) followed by an elastic softening as evident from the drop of $K_0$ value at 78 GPa as shown in Figure 5. This surprising elastic hardening at low pressure range and softening at higher pressure can be attributed to magnetic transitions as observed in the case of reported high pressure phase of Fe$_7$C$_3$ (Nakajima et al., 2011; Chen et al., 2012; Chen et al., 2014; Prescher et al., 2015). Though we have not performed any magnetic measurements on our sample, we propose the anomalies observed in the compression behaviour of $o$-Fe$_7$(C, Si)$_3$ unit cell are probably related to the above magnetic transitions in comparison with literature (Nakajima et al., 2011; Mookherjee et al., 2011; Chen et al., 2012; Prescher et al., 2012; Chen et al., 2014; Prescher et al., 2015). Possibly the pressure induced anisotropic strain induces these magnetic transitions. We have compared our isothermal bulk modulus ($K_0$) extrapolated to inner core pressure at 300 K with that of other phases of $Fe_7C_3$ calculated using EOS fitted parameters and PREM data in the inset of Figure 5 (Dziewonski & Anderson 1981; Nakajima et al., 2011; Chen et al., 2012; Prescher et al., 2015). As evident from the inset of Figure 5, our $K_0$ values approach very close to the PREM data among all having a non-magnetic phase. The value seems to be 8-9% less than that of PREM data in the shown pressure range and is best matched in comparison to other reported values for the non-magnetic phase. Given the density and the bulk modulus values, we propose that $o-Fe_7(C, Si)_3$ can be a strong contender for the composition of Earth's inner core. More measurements are also necessary to resolve the inner core constituent.

## 4 Conclusions

We have synthesized $o$-$Fe_7(C, Si)_3$, followed by characterization using TEM, Raman, and XRD measurements. High-pressure XRD measurements reveal no structural transition up to 104 GPa at room temperature, the highest pressure in this study. High-pressure compression behaviour of our sample reveals two anomalies around 28 and 78 GPa and anisotropic compression of the unit cell. Isothermal bulk modulus value shows elastic stiffening around 28 GPa followed by an elastic softening around 78 GPa. We attribute these anomalies to the magnetic transitions. We have estimated the bulk modulus and the density down to the inner core pressures and temperatures by extrapolating our results. The bulk modulus approaches very close to the PREM data. Also, our estimated density of the inner core at 7000 K seems to have an excellent match with PREM data in comparison to other proposed iron carbides.

**Acknowledgement(s)**

GDM wishes to thank the Ministry of Earth Sciences, Government of India for financial support under the project grant no. MoES/16/25/10-RDEAS. PS wishes to thank DST, INSPIRE program by Department of Science and Technology, Government of India for financial support. We acknowledge DESY (Hamburg, Germany), a member of the Helmholtz Association HGF, for the provision of experimental facilities. Parts of this research were carried out at PETRA III and we would like to thank beamline scientists for assistance in using P02.2 beamline. GDM and PS gratefully acknowledge the financial support from Department of Science and Technology, Government of India for DST-DESY project to carry out the proposed experiment on DESY, Germany.

**Author Contributions**: All authors have equal contribution. All authors reviewed the manuscript.

**References**

Anderson, O. L. (1984) A universal thermal equation-of-state. Journal of Geodynamics, 1, 185-214. https://doi.org/10.1016/0264-3707(84)90027-9

Angel, R. J. (2001) Equation of state. in High-pressure, High-Temperature Crystal Chemistry, Reviews in Mineralology and Geochemistry, 41, pp. 35-60, edited by R. M. Hazen and R. T. Downs, CRC Press, Boka Raton, Fla. https://doi.org/10.2138/rmg.2000.41.2

Birch, F. (1947) Finite Elastic Strain of Cubic Crystals. Physical Review B, 71, 809. https://doi.org/10.1103/PhysRev.71.809

Birch, F. (1952) Elasticity and constitution of the Earth's interior. Journal of Geophysical Research, 57, 227-286. https://doi.org/10.1029/JZ057i002p00227

Boehler, R., von Bargen, N., and Chopelas, A. (1990) Melting, Thermal Expansion, and Phase Transitions of Iron at High Pressures. Journal of Geophysical Research, 95, 21731-21736. https://doi.org/10.1029/JB095iB13p21731

Boehler, R. (1996) Melting temperature of the Earth's mantle and core: Earth's thermal structure. Annual Review of Earth Planetary Sciences, 24, 15-40. https://doi.org/10.1146/annurev.earth.24.1.15

Chen, B., Gao, L., Lavina, B., Dera, P., E. Alp, E., Zhao, J., and Li, J. (2012) Magneto-elastic coupling in compressed $Fe_7C_3$ supports carbon in Earth's inner core. Geophysical Research Letters, 39, L18301. https://doi.org/10.1029/2012GL052875

Chen, B., Li, Z., Zhang, D., Liu, J., Y. Hu, M., Zhao, J., Bi, W., Ercan Alp, E., Xiao, Y., Chow, P. and Li, J. (2014) Hidden carbon in Earth's inner core revealed by shear softening in dense $Fe_7C_3$. PNAS, 111, 17755-17758. https://doi.org/10.1073/pnas.1411154111

Das, T., Chatterjee, S., Ghosh, S., and Saha-Dasgupta, T. (2017) First-principles prediction of Si-doped Fe carbide as one of the possible constituents of Earth's inner core. Geophysical Research Letters, 44, 8776-8784. https://doi.org/10.1002/2017GL073545

Dubrovinsky, L. S., Saxena, S.K., Tutti, F., Rekhi, S., and LeBehan, T. (2000) In Situ X-Ray Study of Thermal Expansion and Phase Transition of Iron at Multimegabar Pressure. Physical Review Letters, 84, 1720-1723. https://doi.org/10.1103/PhysRevLett.84.1720

Dziewonski, A. M., and Anderson, D. L. (1981) Preliminary reference Earth model. Physics of the Earth and Planetary Interiors, 25, 297-356. https://doi.org/10.1016/0031-9201(81)90046-7

Jacobsen, S. D., Holl, C.M., Adams, K.A., Fischer, R.A., Martin, E.S., Bina, C.R., Lin, J.-F., Prakapenka, V.B., Kubo, A. and Dera, P. (2008) Compression of single-crystal magnesium oxide to 118 GPa and a ruby pressure gauge for helium pressure media. American Mineralogist, 93, 1823-1828. https://doi.org/10.2138/am.2008.2988

Li, J., Mao, H.K., Fei, Y., Greoryanz, E., Eremets, M., and Zha, C.S. (2002) Compression of $Fe_3C$ to 30 GPa at room temperature. Physics and Chemistry of Minerals, 29, 166-169. https://doi.org/10.1007/s00269-001-0224-4

Li, J., and Fei, Y. (2003) Experimental constraints on core composition. in Treatise on Geochemistry (2nd edition), edited by H. D. Holland and K. K. Turekian, pp. 527-557, Elsevier, Oxford. https://doi.org/10.1016/B0-08-043751-6/02014-4

Lin. K.-L. (2014) Phase Identification Using Series of Selected Area Diffraction Patterns and Energy Dispersive Spectrometry within TEM. Microscopy Research, 2, 57-66. http://dx.doi.org/10.4236/mr.2014.24008

Liu, J., Li, J., and Ikuta, D. (2016) Elastic softening in $Fe_7C_3$ with implications for Earth's deep carbon reservoirs. Journal of Geophysical Research: Solid Earth, 121, 1514-1524. https://doi.org/10.1002/2015JB012701

Mao, H. K., Wu, Y., Chen, L.C., and Shu, J.F. (1990) Static Compression of Iron to 300 GPa and $Fe_{0.8}Ni_{0.2}$ Alloy to 260 GPa: Implications for Composition of the Core. Journal of Geophysical Research, 95, 21737-21742. https://doi.org/10.1029/JB095iB13p21737

Mookherjee, M., Nakajima, Y., Steinle-Neumann, G., Glazyrin, K., Wu, X., Dubrovinsky, L., McCammon, C., Chumakov, A. (2011) High-pressure behavior of iron carbide ($Fe_7C_3$) at inner core conditions. Journal of Geophysical Research, 116, B04201. https://doi.org/10.1029/2010JB007819

Mao, H. K., Xu, J., Bell, P.M. (1986) Calibration of the Ruby pressure Gauge to 800 kbar Under Quasi-Hydrostatic Conditions. Journal of Geophysical Research, 91, 4673-4676. https://doi.org/10.1029/JB091iB05p04673

Mukherjee, G. D. & Boehler, R. (2007) High-pressure Melting Curve of Nitrogen and the Liquid-Liquid Phase Transition. Physical Review Letters, 99, 225701. https://doi.org/10.1103/PhysRevLett.99.225701

Murnaghan, F. D. (1937) FINITE DEFORMATIONS OF AN ELASTIC SOLID. American Journal of Mathematics, 59, 235. https://doi.org/10.2307/2371405

Nakajima, Y., Takahashi, E., Suzuki, T., and Funakoshi, F. (2009) "Carbon at the core" revisited. Physics of the Earth and Planetary Interiors, 174, 202-211. https://doi.org/10.1016/j.pepi.2008.05.014

Nakajima, Y., Takahashi, E., Sata, N., Nishihara, Y., Hirose, K., Funakoshi, F, and Ohishi, Y. (2011) Thermoelastic property and high-pressure stability of $Fe_7C_3$: Implication for iron-carbide in the Earth's core. American Mineralogist, 96, 1158-1165. https://doi.org/10.2138/am.2011.3703

Poirier, J. P., 1994. Light-elements in the Earth's outer core. Physics of the Earth and Planetary Interiors, 85, 319-337. https://doi.org/10.1016/0031-9201(94)90120-1

Polian, A., Gauthier, M., Souza, S.M., Triches, D.M., Cardoso de Lima, J., Antonio Grandi, T. (2011) Two-dimensional pressure-induced electronic topological transition in $Bi_2Te_3$. Physical Review B 83, 113106. https://doi.org/10.1103/PhysRevB.83.113106

Prescher, C., Dubrovinsky, L., McCammon, C., Glazyrin, K., Nakajima, Y., Kantor, A., Merlini, M., and Hanflan, M. (2012) Structurally hidden magnetic transitions in $Fe_3C$ at high pressures. Physical Review B, 85, 140402(R). https://doi.org/10.1103/PhysRevB.85.140402

Prescher, C., Dubrovinsky, L., Bykova, E., Kupenko, I., Glazyrin, K., McCammon, C., Mookherjee, M., Nakajima, Y., Miyajima, N.,Sinmyo, R., Cerantola, V., Dubrovinskaia, N., Prakapenka, V., Ruffer, R., Chumakov, A., and Hanfland, M. (2015) High Poisson's ratio of Earth's inner core explained by carbon alloying. Nature Geoscience, 8, 220. https://doi.org/10.1038/ngeo2370

Prescher, C., and Prakapenka, V. B., (2015b) DIOPTAS: a program for reduction of two-dimensional X-ray diffraction data and data exploration. High Pressure Research, 35 223-230. https://doi.org/10.1080/08957959.2015.1059835

Planck, M. (1901) On the Law of the Energy Distribution in the Normal Spectrum. Annalen der Physik, 4, 553.

Saha, P., and Mukherjee, G. D. (2020) Temperature measurement in double sided Laser Heated Diamond Anvil Cell and reaction of carbon. Indian Journal of Physics., 1-8. https://doi.org/10.1007/s12648-020-01699-2.

Saha, P., Mazumdar, A., and Mukherjee, G.D. (2020) Thermal conductivity of dense hcp iron: direct measurements using laser heated diamond anvil cell. Geoscience Frontiers., 11, 1755-1761, https://doi.org/10.1016/j.gsf.2019.12.010.

Sata, N., Hirose, K., Shen, G., Najima, Y., Ohishi, Y., Hirao, N. (2010) Compression of FeSi, $Fe_3C$, $Fe_{0.95}O$, and FeS under the core pressures and implication for light element in the Earth's core. Journal of Geophysical Research, 115, B09204. https://doi.org/10.1029/2009JB006975

Stevenson, D. J. (1981) Models of Earth's core. Science, 214, 611-619. https://doi.org/10.1126/science.214.4521.611

Shirley, R. (2002) The CRYSFIRE 2002 System for Automatic Powder Indexing: Users Manual (TheLattice Press, Guildford).

Toby, B. H. (2001) EXPGUI, a graphical user interface for GSAS. Journal of Applied Crystallography, 34, 210. https://dx.doi.org/10.1107/S0021889801002242

Vočadlo, L., Brodholt, J., P Dobson, D., Knight, K.S., Marshall, W.G., David Price, G., G Wood, I. (2002) The effect of the ferromagnetism on the equation of state of $Fe_3C$ studied by first principles calculations. Earth and Planetary Science Letters, 203, 567-575. https://doi.org/10.1016/S0012-821X(02)00839-7

Wood, B. J. (1993) Carbon in the core. Earth and Planetary Science Letters, 117, 593-607. https://doi.org/10.1016/0012-821X(93)90105-I

**Table-1:** High-pressure high-temperature synthesis conditions, experimental stations, obtained lattice parameters. Synthesis pressure shown below is measured before heating at 300 K.

| Station | In-house | In-house | In-house | P02.2 beamline* Run #1 | P02.2 beamline Run#2 | P02.2 beamline Run#3 | P02.2 beamline Run#4 |
|---|---|---|---|---|---|---|---|
| Synthesis P (GPa) & T (K) | 15 ~2000 | 28 ~2000 | 8 ~2000 | 8.5 ~2250 | 20.5 ~2300 | 18 ~2175 | 35 ~2080 |
| PTM | NaCl | NaCl | NaCl | MgO | MgO | MgO | MgO |
| Analysis Condition | recovered | recovered | recovered | 9.4 GPa | 23.4 GPa | 20.6 GPa | 49 GPa |
| $R_{wp}$ | 1.8% | 2% | 3.5% | 1% | 1.1% | 1% | 0.9% |
| Data shown | Figure S3(a) | Figure S3(b) | Figure S3(c) | Figure S6(a) | Figure S7(a) | Figure S8(a) | Figure 1(a) |
| $a$ (Å) | 12.476(2) | 12.479(2) | 12.406(2) | 12.119(1) | 11.629(1) | 11.775(3) | 11.329(3) |
| $b$ (Å) | 4.505(1) | 4.488(1) | 4.506(1) | 4.403(2) | 4.336(2) | 4.359(2) | 4.231(2) |
| $c$ (Å) | 14.597(2) | 14.613(2) | 14.630(3) | 13.940(1) | 13.414(1) | 13.506(4) | 13.245(5) |
| $V$ (Å$^3$) | 820.39(12) | 818.49(16) | 818.30(12) | 743.76(10) | 676.27(9) | 693.20(4) | 634.90(8) |

* Synthesis carried out using LHDAC set-up facility in P02.2 beamline at Petra-III, DESY. In each run, after synthesis, pressure is increased to a maximum achievable value and then released.

**Table-2:** Comparison of the observed 2θ values with those of hexagonal (Nakajima *et al.*, 2011) and orthorhombic phases at three different pressure points in this study. Pressures are taken from MgO scale.

| Pressure (GPa) Run No. | $2\theta_{obs}$ | $2\theta_{hexa}$ | $2\theta_{obs} - 2\theta_{hexa}$ | $(hlk)_{hexa}$ | $2\theta_{orth}$ | $2\theta_{obs} - 2\theta_{orth}$ | $(hlk)_{orth}$ |
|---|---|---|---|---|---|---|---|
| 9.4 Run-1 9.9 (Nakajima *et al.*, 2011) | 8.21 | | | | 8.21 | 0 | 116 |
| | 8.40 | 8.39 | 0.01 | 121 | 8.39 | 0.01 | 504 |
| | 8.57 | 8.49 | 0.08 | 300 | 8.57 | 0 | 216 |
| | 8.82 | 9.00 | -0.18 | 112 | 8.82 | 0 | 223 |
| | 9.17 | 9.30 | -0.13 | 301 | 9.16 | 0.01 | 611 |
| | 9.60 | 9.45 | 0.15 | 022 | 9.59 | 0.01 | 217 |
| | | 9.83 | -0.23 | 220 | | | |
| | 11.27 | | | | 11.27 | 0 | 621 |
| | 11.87 | | | | 11.86 | 0.01 | 525 |
| 49 Run-4 ~49 (Mean of 53.2 and 44.6, mean lattice parameters are used) (Nakajima *et al.*, 2011) | 7.18 | | | | 7.19 | -0.01 | 411 |
| | 7.71 | 7.74 | -0.03 | 120 | 7.74 | -0.03 | 404 |
| | 8.42 | 8.37 | 0.05 | 012 | 8.44 | -0.02 | 511 |
| | 8.61 | 8.68 | -0.07 | 121 | 8.64 | -0.03 | 116 |
| | 9.36 | 9.34 | 0.02 | 112 | 9.37 | -0.01 | 322 |
| | 9.55 | 9.63 | -0.08 | 301 | 9.53 | 0.02 | 610 |
| | 9.72 | 9.80 | 0.08 | 022 | 9.71 | 0.01 | 117 |
| | 10.03 | 10.15 | -0.12 | 220 | 10.04 | -0.01 | 217 |
| | 11.77 | | | | 11.80 | -0.03 | 621 |
| 72 Run-4 71.5 (Nakajima *et al.*, 2011) | 7.77 | 7.86 | -0.09 | 120 | 7.81 | -0.04 | 404 |
| | 8.53 | 8.62 | -0.09 | 012 | 8.53 | 0 | 511 |
| | 8.72 | 8.84 | -0.12 | 121 | 8.73 | -0.01 | 116 |
| | 9.36 | | | | 9.36 | 0 | 223 |
| | 9.62 | 9.60 | 0.02 | 112 | 9.65 | -0.03 | 406 |

|  | 9.79 | 9.80 | 0.01 | 301 | 9.78 | 0.01 | 610 |
|  | 9.88 |  |  |  | 9.86 | -0.02 | 117 |
|  | 10.16 | 10.05 | 0.11 | 022 | 10.19 | -0.03 | 217 |
|  | 12.06 |  |  |  | 12.06 | 0 | 621 |

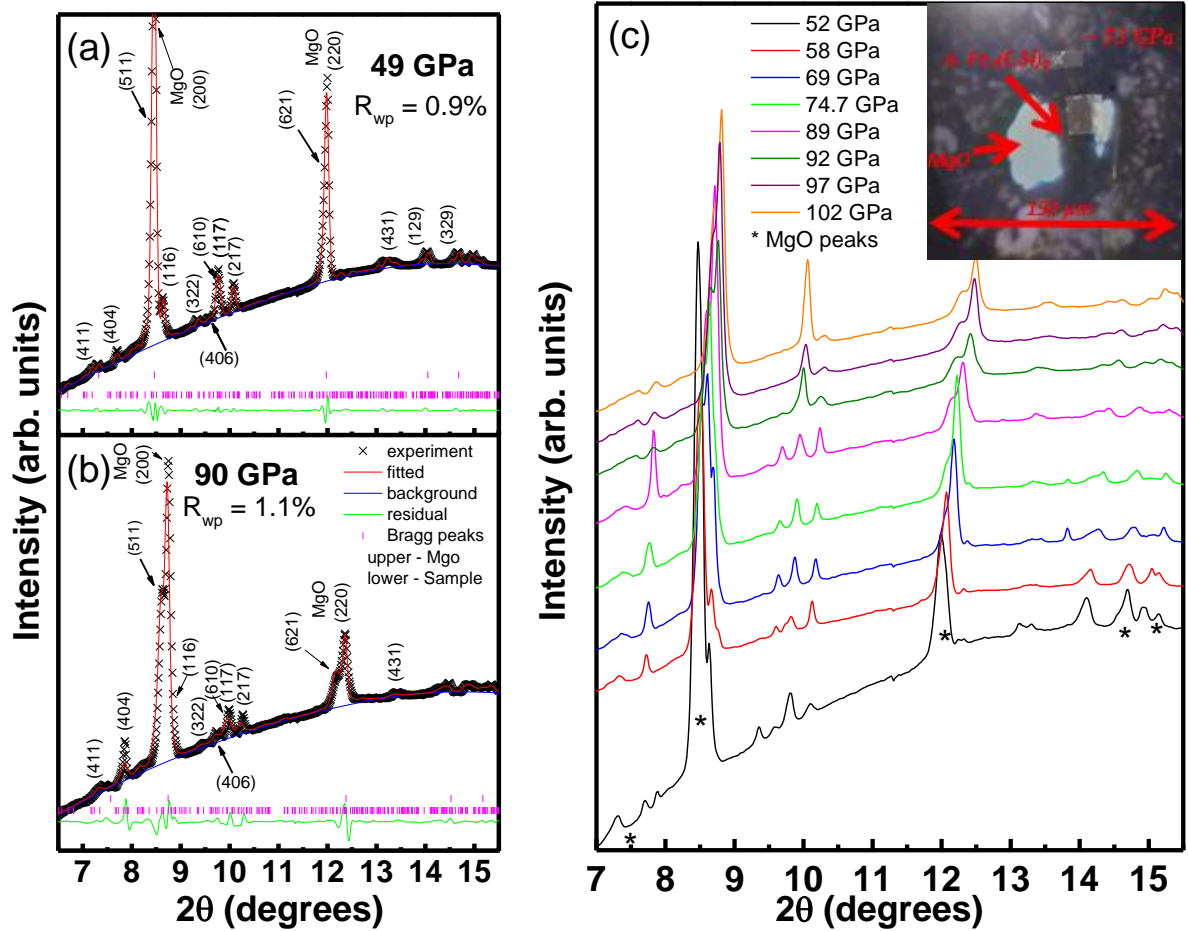

**Figure 1.** X-ray diffraction pattern of the synthesized sample in Run-4 at (a) 49 GPa, and (b) 90 GPa with their Le-bail fitting using GSAS (Toby, 2001). (c) Evolution of diffraction patterns at selected pressure values at 300 K up to 102 GPa of $o$-Fe$_7$(C, Si)$_3$ collected in Run-4. Inset: Image of loaded LHDAC around 53 GPa containing synthesized $o$-Fe$_7$(C, Si)$_3$ and MgO as pressure transmitting medium in the central hole of diameter around 65 μm.

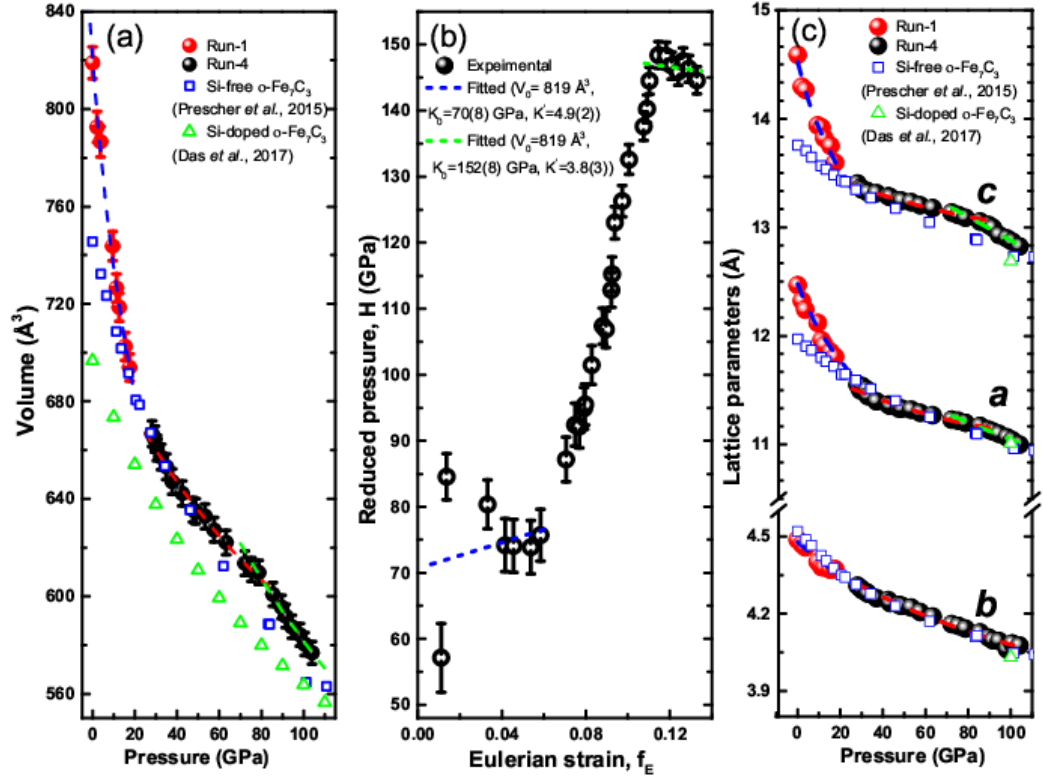

**Figure 2.** (a) Comparison of the compression behaviour of $o$-Fe$_7$(C, Si)$_3$ at 300 K up to 104 GPa (all filled symbols) with Si-free (blue opened square symbols) and Si-doped (green opened triangular symbols) $o$-Fe$_7$C$_3$. Red filled circles, and black filled circles are from Run-1 and Run-4, respectively. Dashed blue, red, and green curves represent 3$^{rd}$-order B-M EOS fitting for three distinct ranges of pressure. (b) The reduced pressure *vs.* the finite Eulerian strain reveals change in the slope at two pressure points in its linear behavior around 28 GPa and 78 GPa. Blue and green dotted lines represent the fitting to Eqn. $H = K_0 + 3/2K_0(K' - 4)f_E$. (c) Axial compression behaviour of $o$-Fe$_7$(C, Si)$_3$ (red, and black filled circle from Run-1, and Run-4, respectively) at 300 K along with Si-free (blue opened square symbols) and Si-doped (green opened triangular symbols) $o$-Fe$_7$C$_3$ as a function of pressure. Blue, red, and green dashed lines are fitting to 3$^{rd}$-order B−M EOS of our data at three different ranges of pressure.

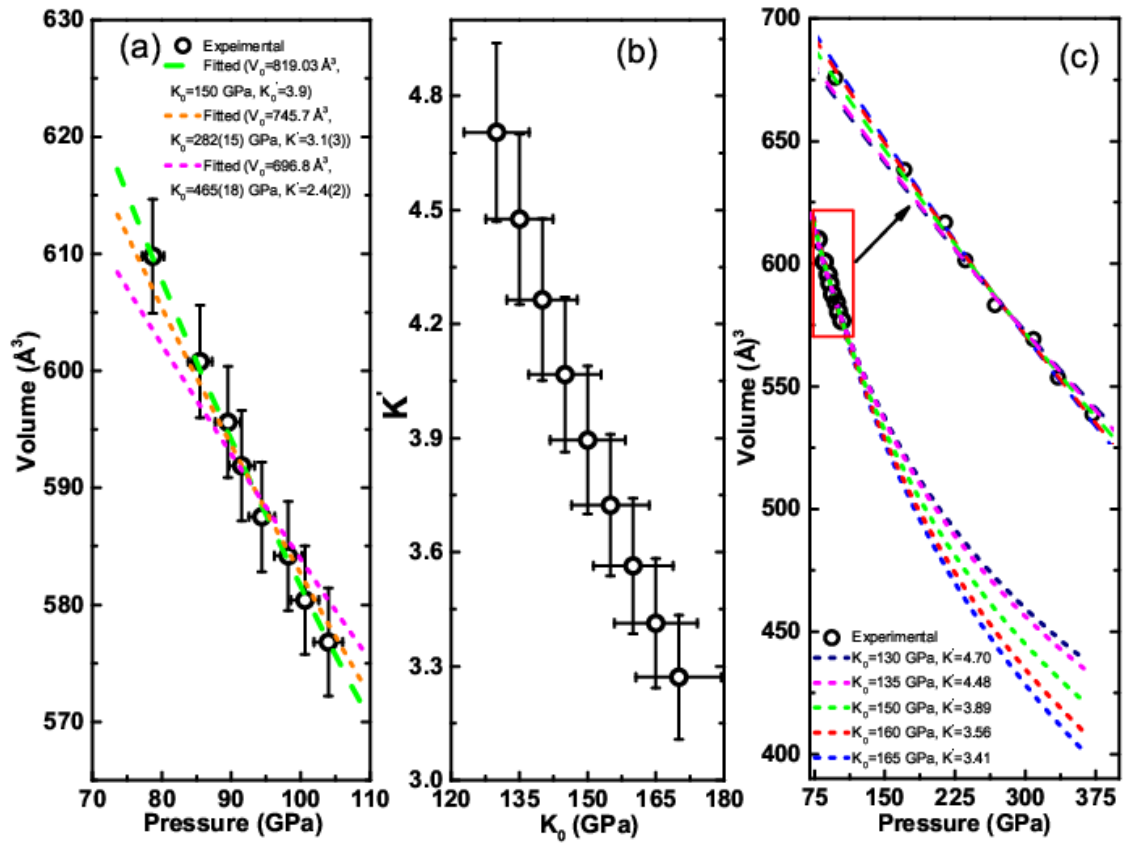

**Figure 3.** (a) Comparison of the fitting of the data in range (iii), taking different $V_0$ values. Green dashed curve represents fitted curve with $V_0$ of this study, orange dotted curve represents fitted curve with $V_0$ taken from Prescher *et al.* (2015) of pure $o$-$Fe_7C_3$, and magenta dotted curve represents fitted curve with $V_0$ taken from Das *et al.* (2017) of Si-doped $o$-$Fe_7C_3$. (b) Correlation of $K_0$, and $K'$ for the fitting of the data in range (iii), taking $V_0$ from this study. (c) A few fitting curves by varying $K_0$, and $K'$ extrapolated to Earth's core pressure 360 GPa. Inset is the zoomed-in view of the red rectangle region.

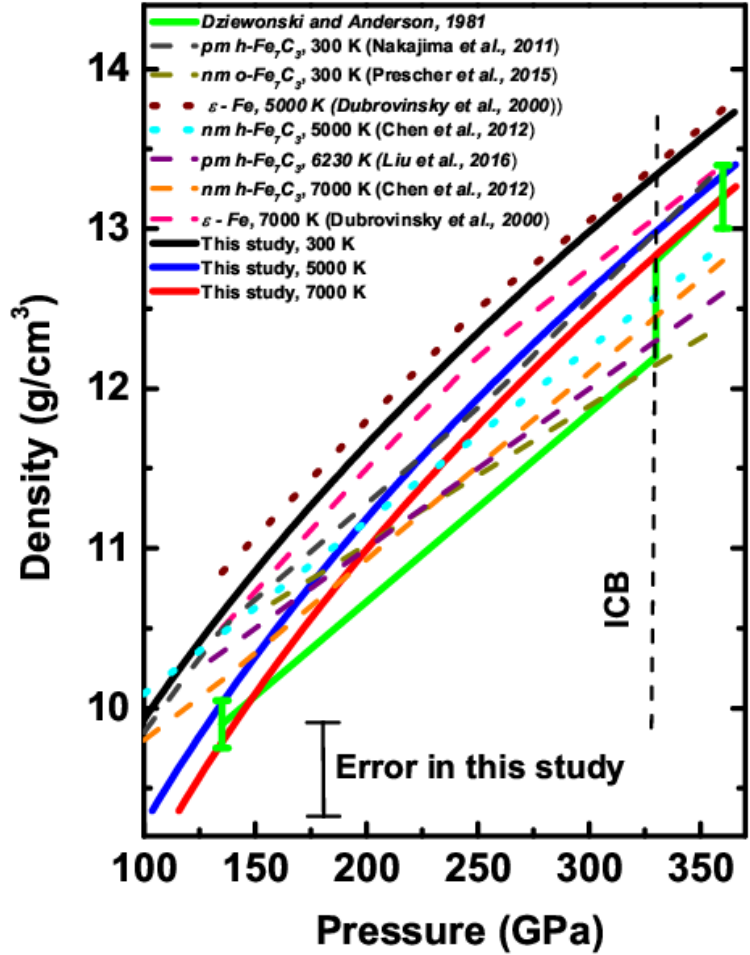

**Figure 4.** Comparison of densities of $\varepsilon-$Fe, $Fe_7C_3$, and $o$-$Fe_7(C, Si)_3$ at the Earth's core pressures. Green curve is the density profile of PREM (Dziewonski & Anderson, 1981). Wine dotted and pink dashed curves represent isothermal density profile of $\varepsilon-$Fe at temperatures 5000 K and 7000 K (Dubrovinsky *et al.*, 2000). Dark grey dashed and purple dashed curves show density profile of pm-$h-Fe_7C_3$ at 300 K (Nakajima *et al.*, 2011) and 6230 K (Liu *et al.*, 2016). Cyan dotted and orange dashed curves are the density profile of nm-$h-Fe_7C_3$ at 5000 K and 7000 K (Chen *et al.*, 2012). Dark yellow dashed curve represents density profile of nm $o-Fe_7C_3$ at 300 K (Prescher *et al.*, 2015). Black, blue, and red represent the density profile at 300 K, 5000 K, and 7000 K, respectively, in our study obtained by extrapolating the density of range (iii). Black vertical line represents the error in density measurements.

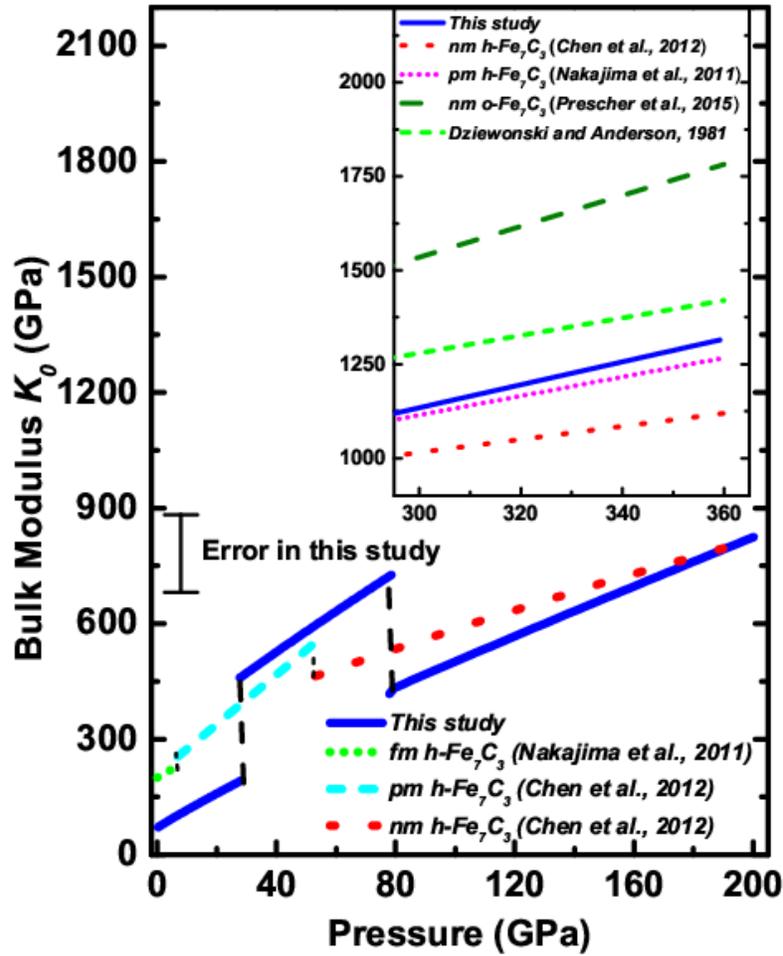

**Figure 5.** $K_0$ values of $o$-Fe$_7$(C, Si)$_3$ at 300 K as a function of pressure, estimated by extrapolating the parameters obtained from fitted 3$^{rd}$-order B−M EOS in the range (iii) of our study (blue lines), ferromagnetic (fm)- phase of $h$−Fe$_7$C$_3$ (green short dashed line, Nakajima *et al.*, (2011)), paramagnetic(pm)-phase of $h$−Fe$_7$C$_3$ (cyan dotted line, Chen *et al.*, (2012)), and nonmagnetic (nm)-phase of $h$−Fe$_7$C$_3$ (red dotted line, Chen *et al.*, (2012)). Inset: $K_0$ extrapolated to core pressure at 300 K of our study (blue line), pm-phase of $h$−Fe$_7$C$_3$ (cyan short dotted line, Nakajima *et al.*, (2011)), nm-phase of $h$−Fe$_7$C$_3$ (red dotted line, Chen et al.*,* (2012)), nm-phase of $o$−Fe$_7$C$_3$ (olive dashed line, Prescher *et al.*, (2015)), PREM data (Magenta short dashed line, Dziewonski & Anderson (1981)).

Supplementary Information for

**Synthesis and Compression study of orthorhombic $Fe_7(C, Si)_3$: A possible constituent of the Earth's core**

Pinku Saha[1], Konstantin Glazyrin[2], and Goutam Dev Mukherjee[1*]

[1]*National Centre for High Pressure Studies, Department of Physical Sciences, Indian Institute of Science Education and Research Kolkata, Mohanpur Campus, Mohanpur 741246, Nadia, West Bengal, India.*

[2]*Photon Science, Deutsches Elektronen Synchrotron, 22607 Hamburg, Germany*

In this Supplementary information, figures and tables are provided in the support of the main manuscript. We have carried out scanning electron microscopic (SEM) imaging and elemental mapping on the starting mixture, which are shown in Figure S1. Elemental mapping and spectra measurements are carried out from 4×3 μm$^2$ rectangular cross section shown by magenta rectangle in the electron image, which reveal the presence of the Fe, C, and Si in the starting mixture. Raman spectra of the unheated and the heated recovered samples at three different runs are compared in Figure S2. Three prominent modes are observed in the unheated sample at around 520, 1330 and 1580 cm$^{-1}$ corresponding to Silicon and D-band and G-band of nano-diamond. But the Raman spectra of the heated sample do not show any prominent peak of the parent constituents, which excludes the presence of un-reacted elemental C and Si in the heated samples. XRD patterns of the retrieved samples of *in-house* experiments along with their Le-bail fitting are shown in Figure S3. Figure S4 represent the image of mapping of the chemical composition of the retrieved samples at different runs (Run-3, and 4) carried out at P02.2 beamline within an area ranging from 300 nm$^2$ to 3 μm$^2$, which reveal that our retrieved sample contains iron, carbon, and silicon. The high percentage of the carbon is evident from the image, which is due to the use of carbon coated copper grid during TEM measurements. The corrected percentage of Fe, C, and Si in the synthesized compound is found to be at atomic ratio Fe : C : Si = 70(±2) : 25(±2) : 5(±1), which is very close to the starting materials. The Selected area diffraction pattern (SADP) is shown in Figure S5. Figure S6-S8 represent XRD patterns with their Le-bail fitting, and the pattern evolutions at selected pressure points of Run# 1-3, respectively. Comparison of experimental patterns with those of computed for hexagonal and orthorhombic phases at three different pressure points are presented in Figure S9. Volume evolution with pressure and their fitting to $3^{rd}$-order B-M EOS at different runs are shown in Figure S10. We have performed XRD measurements from different positions of the synthesized samples at different runs (Run-2 & Run-4) using a pre-selected grid pattern and are shown in Figure S11. A typical view of the positions from where XRD data are collected during grid run is shown in Figure S11 (d). The patterns show no change in Bragg peak positions in each pressure point. All the estimated lattice parameters and volumes lie within error bars. In Figure S12, we have shown the effect of temperature on the pressure of the sample that acts as negative pressure. In Table-S1, the axial compressibility for each axis is listed for three different pressure ranges.

**Table-S1:** The values of linear compressibility (β) of the three lattice parameters in three different regions of the axial compression of $o$-Fe$_7$(C, Si)$_3$. β is represented in units of GPa$^{-1}$.

| Pressure range | Ambient to 28 GPa | 28 to 78 GPa | 78 to 104 GPa |
|---|---|---|---|
| $\beta^a$ | 0.00421(3) | 0.00262(2) | 0.00099(2) |
| $\beta^b$ | 0.00192(2) | 0.00166(3) | 0.00166(3) |
| $\beta^c$ | 0.00558(3) | 0.00098(2) | 0.00104(2) |

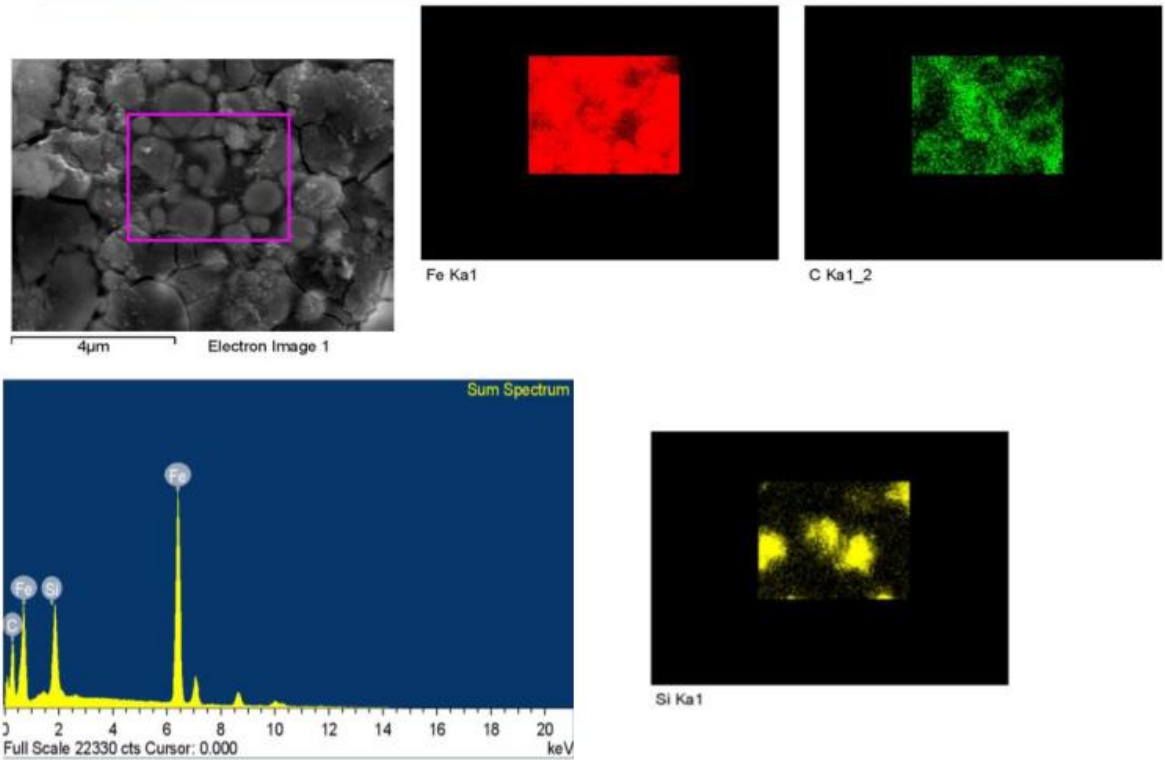

**Figure S1:** Scanning electron microscopy (SEM) imaging and elemental mapping of the starting mixture.

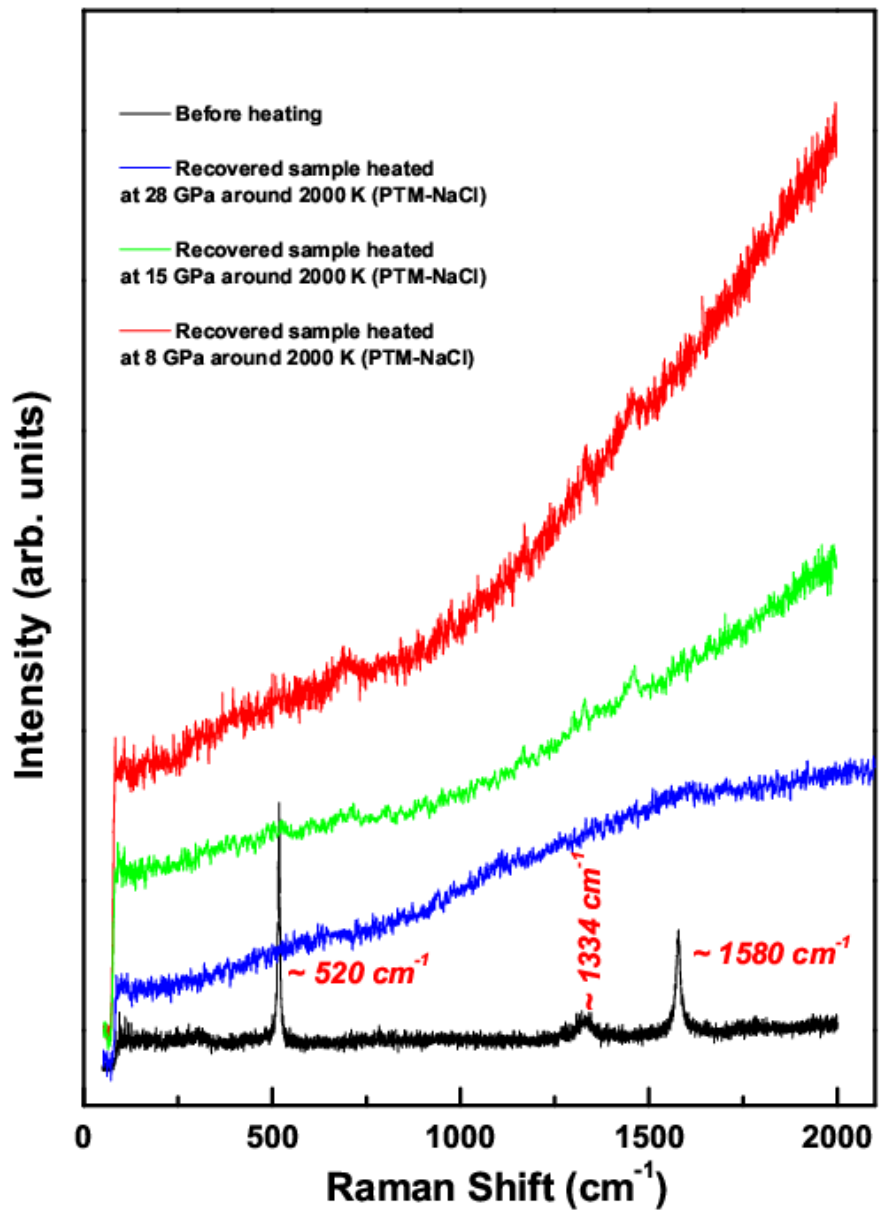

**Figure S2:** Raman spectra of unheated and heat quenched sample at three different pressure points. Heating of the samples are carried out around 2000 K.

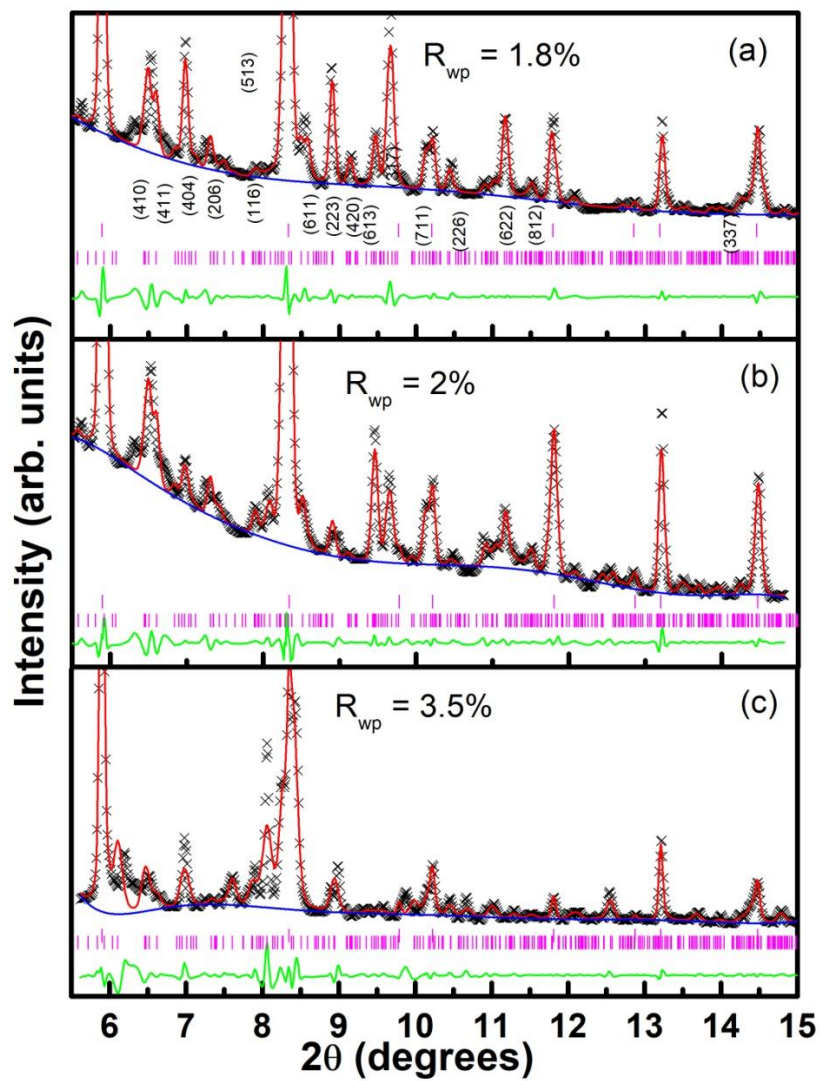

**Figure S3:** XRD patterns with their Le-bail fitting of the synthesized samples retrieved from *in-house* experiments. Heatings are performed at: (a) 15 GPa, (b) 28 GPa, and (c) 8 GPa around 2000 K.

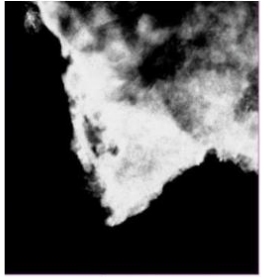
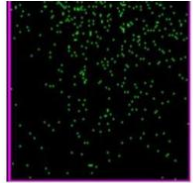
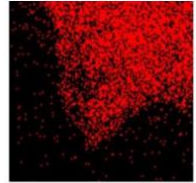
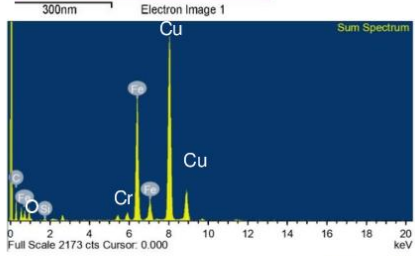
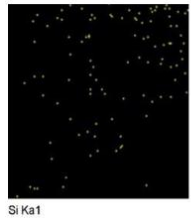

| Element | Atomic % |
|---|---|
| Fe | 71 |
| C | 26 |
| Si | 4 |

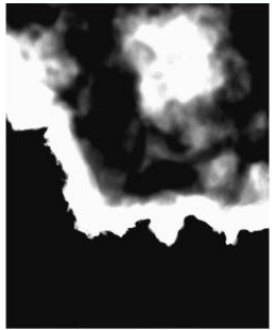
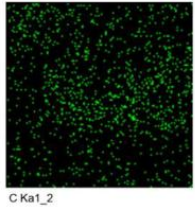
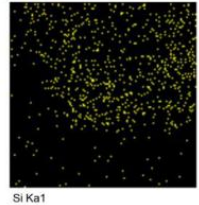
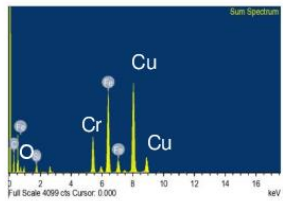
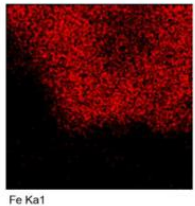

| Element | Atomic % |
|---|---|
| Fe | 71.7 |
| C | 24 |
| Si | 4.3 |

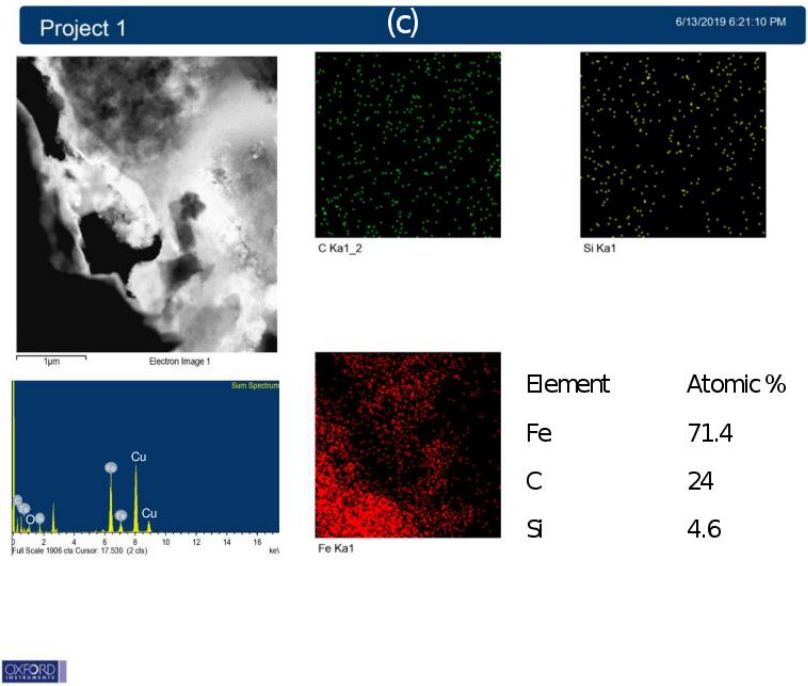

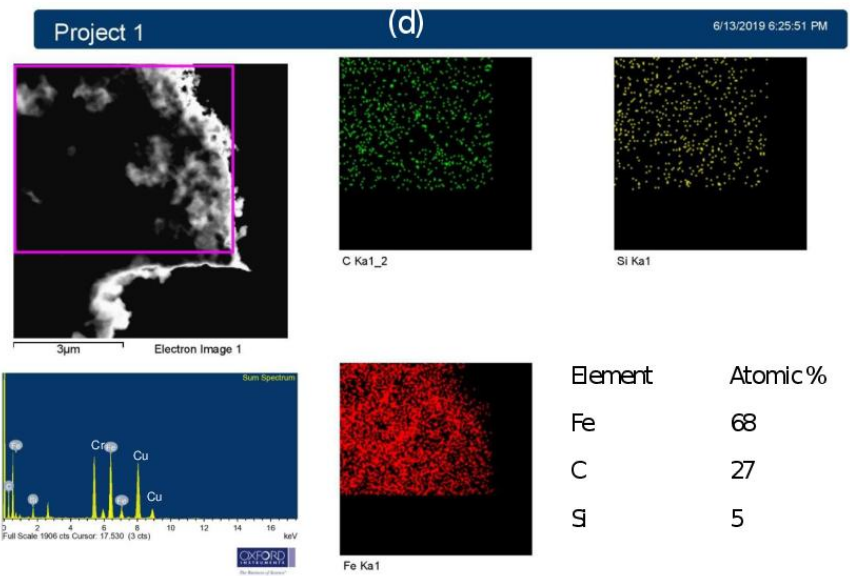

**Figure S4:** The image of mapping of the chemical compositions at different positions on the recovered samples of Run-3 & Run-4 heated in P02.2 beamline around 2000 K taken using TEM operating at 200 kV. (a) & (b) are recovered from Run-4, and (c) & (d) are from Run-3.

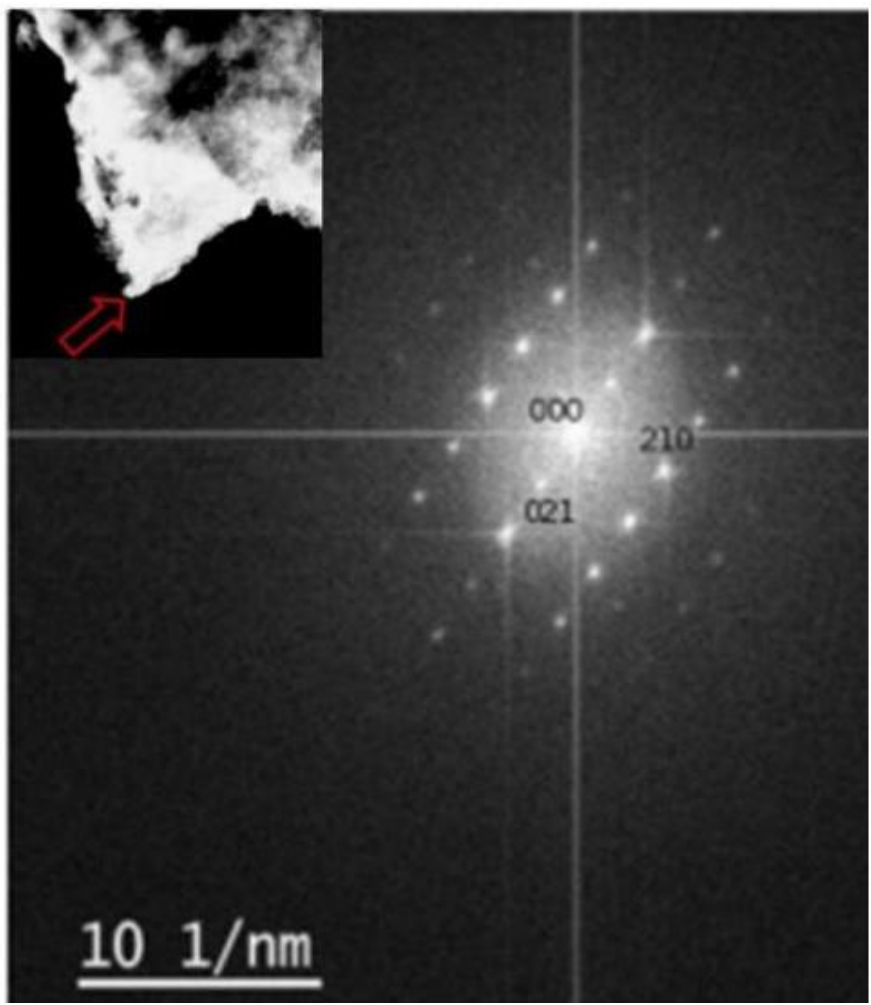

**Figure S5:** The Selected area diffraction pattern (SADP) of the grain (recovered from Run-4) at the tip of red arrow shown in inset. The pattern is indexed to the *Pbca* space group of orthorhombic phase with unit cell parameters a = 12.9(1) Å, b = 4.9(1) Å, and c = 14.2(2) Å.

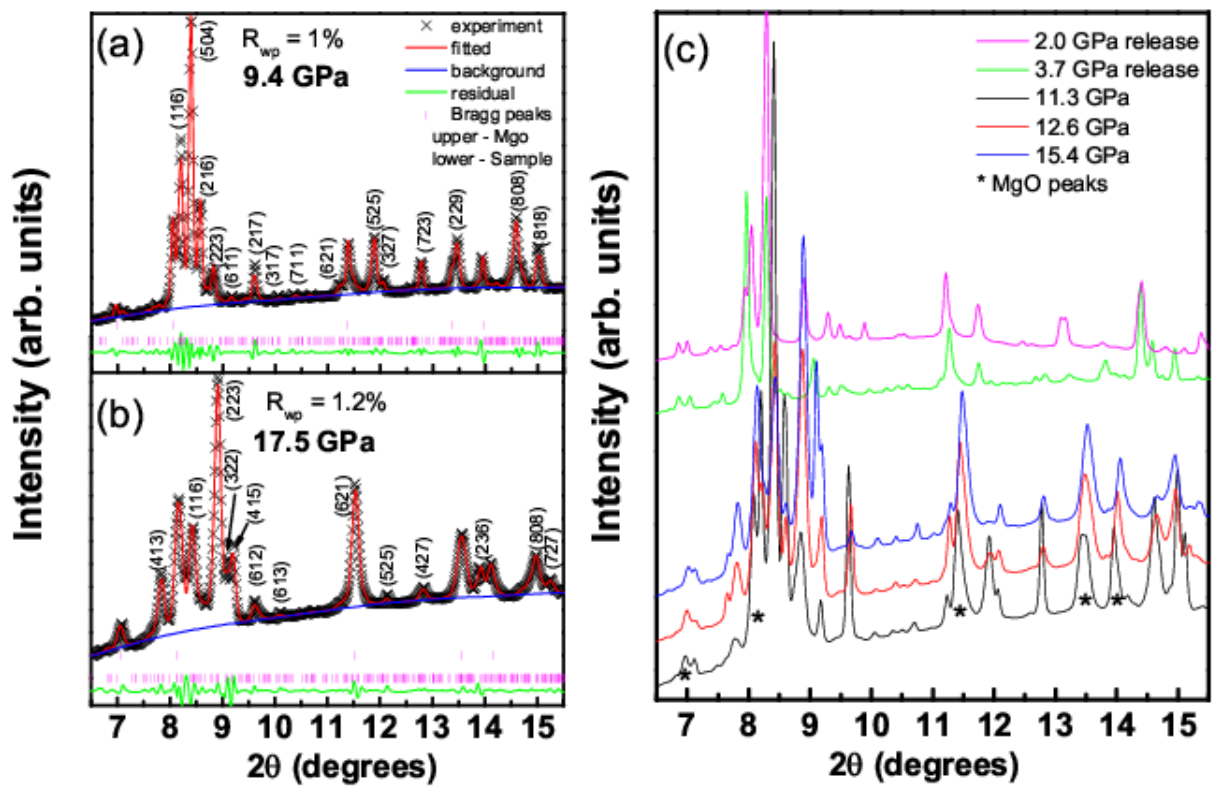

**Figure S6:** X-ray diffraction patterns of the synthesized sample in Run-1 at (a) 9.4 GPa, and (b) 17.5 GPa with their Le-bail fitting. (c) Evolution of selected diffraction patterns with pressure at 300 K. The maximum pressure in this run is 17.5 GPa.

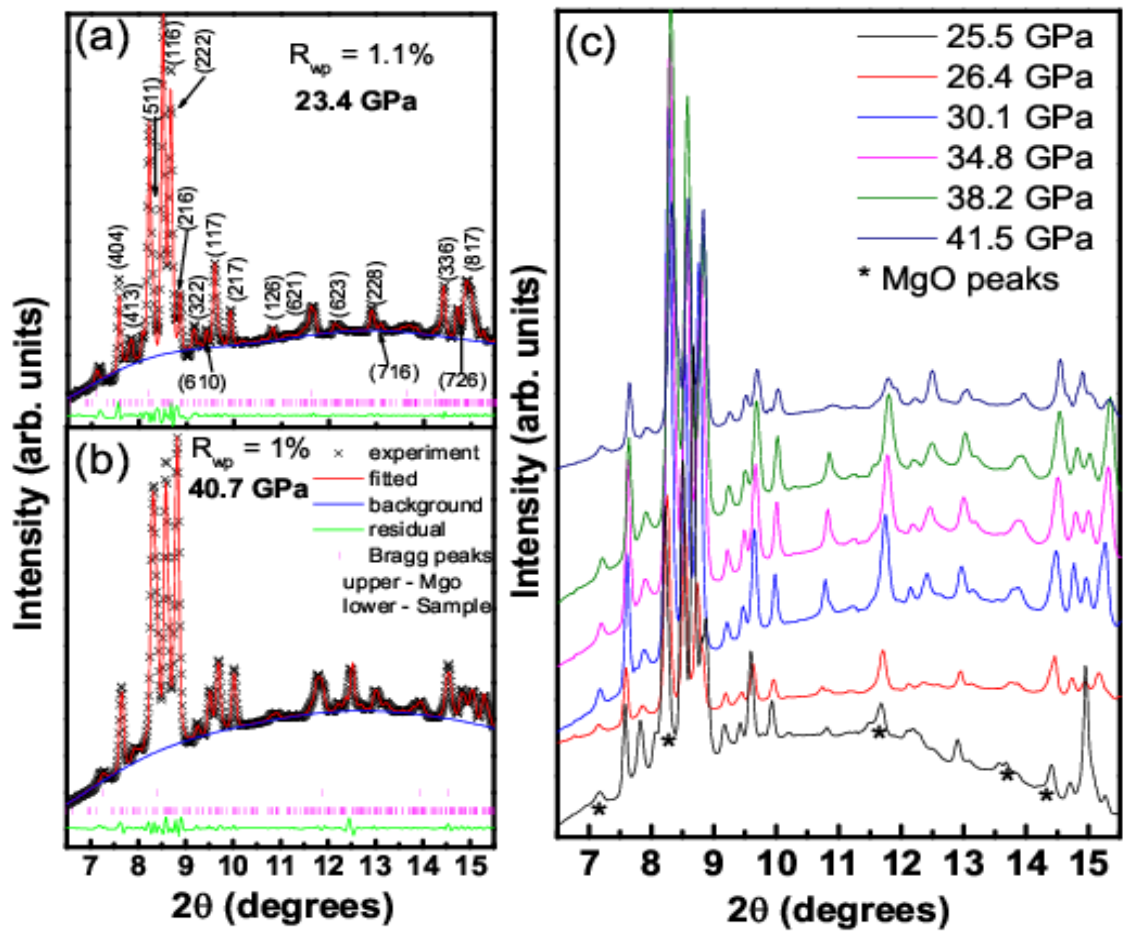

**Figure S7:** X-ray diffraction patterns of the synthesized sample from Run-2 at (a) 23.4 GPa, and (b) 40.7 GPa with their Le-bail fitting. (c) Evolution of selected diffraction patterns with pressure at 300 K. The maximum pressure in this run is 41.5 GPa.

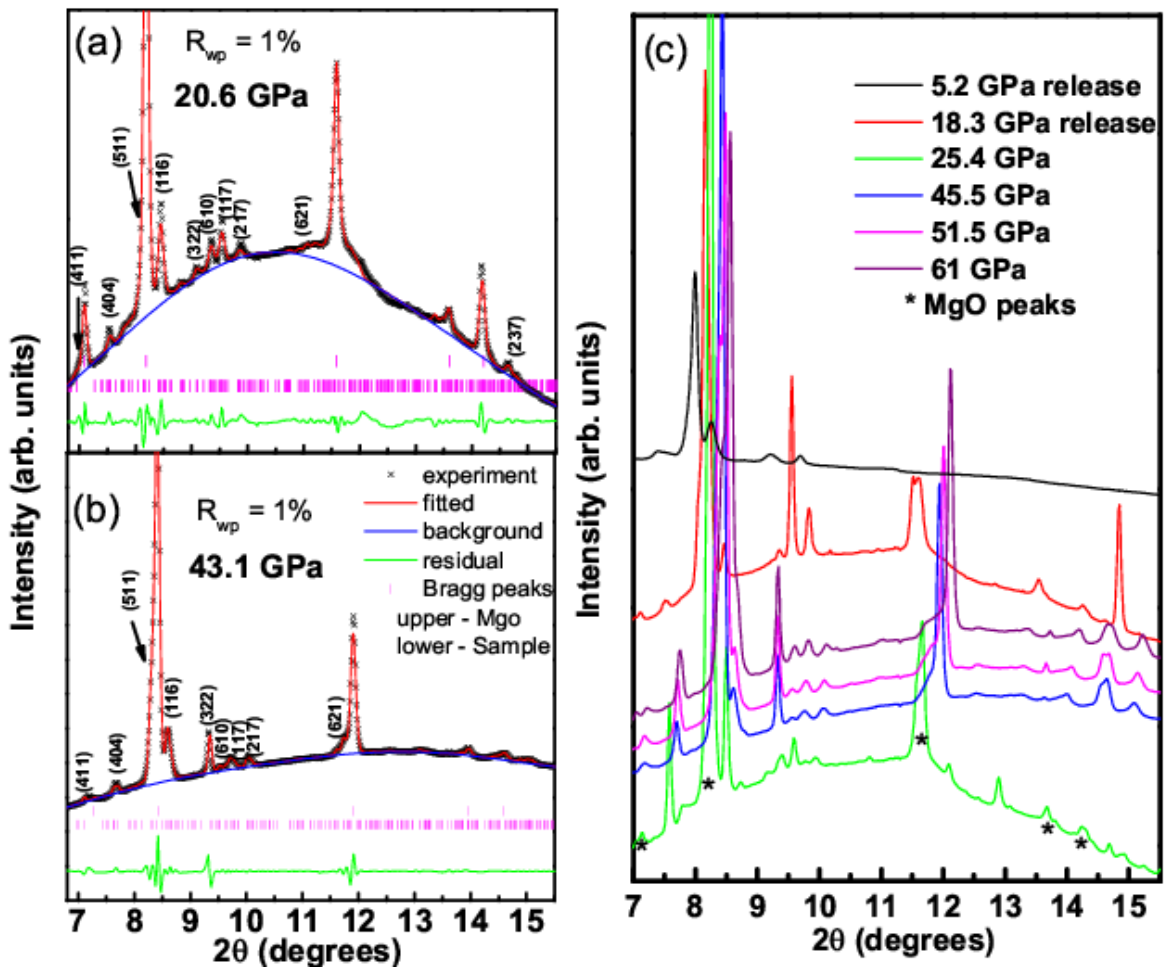

**Figure S8:** X-ray diffraction patterns of the synthesized sample in Run-3 at (a) 20.6 GPa, and (b) 43.1 GPa with their Le-bail fitting. (c) Evolution of selected diffraction patterns with pressure at 300 K. The maximum pressure in this run is 62.5 GPa.

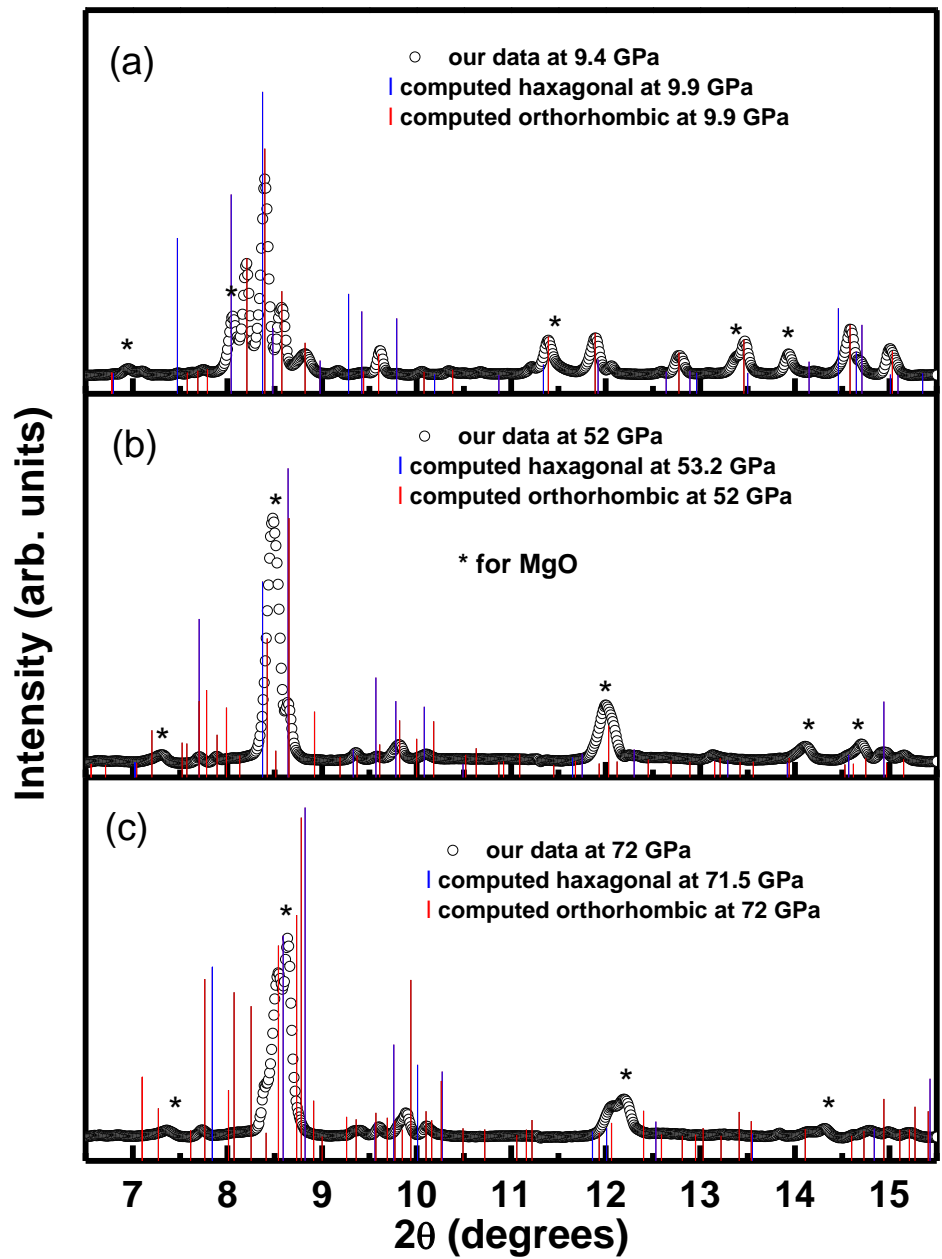

**Figure S9**: Comparison of experimental patterns with those of computed for hexagonal and orthorhombic phases at three different pressure points.

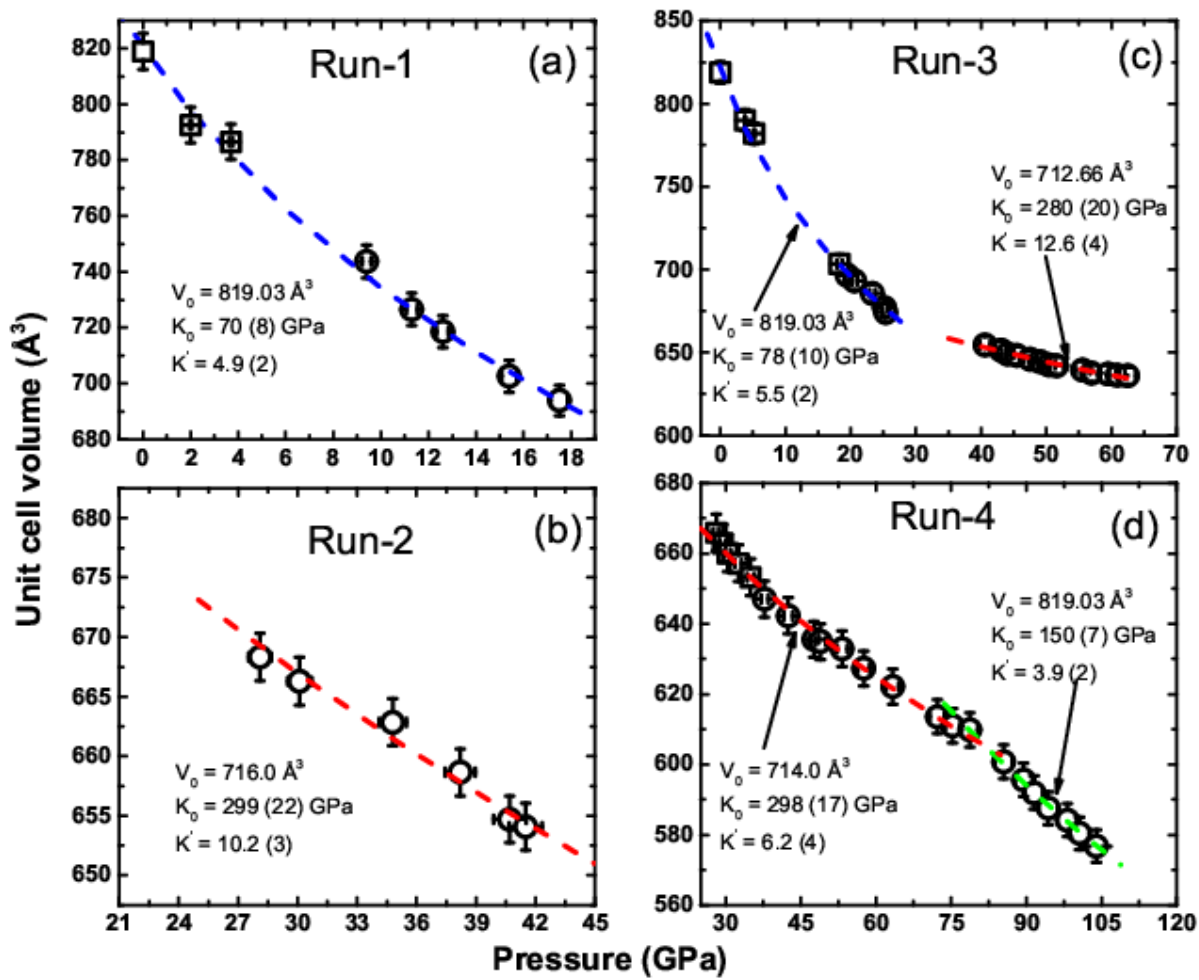

**Figure S10:** Volume evolution of the synthesized samples with pressure in different runs. Open circles represent data points during pressure-increase in each run. Square symbols are data points obtained during release of pressure. The ambient volume of value about ~819 Å$^3$ is taken from the analysis of the XRD patterns of the samples retrieved from *in-house* experiments, and are also represented by square symbol. Dashed lines through data points represent the fitting to 3$^{rd}$-order B-M EOS indifferent ranges (blue dashed – range (i), red dashed – range (ii), and green dashed – range (iii)).

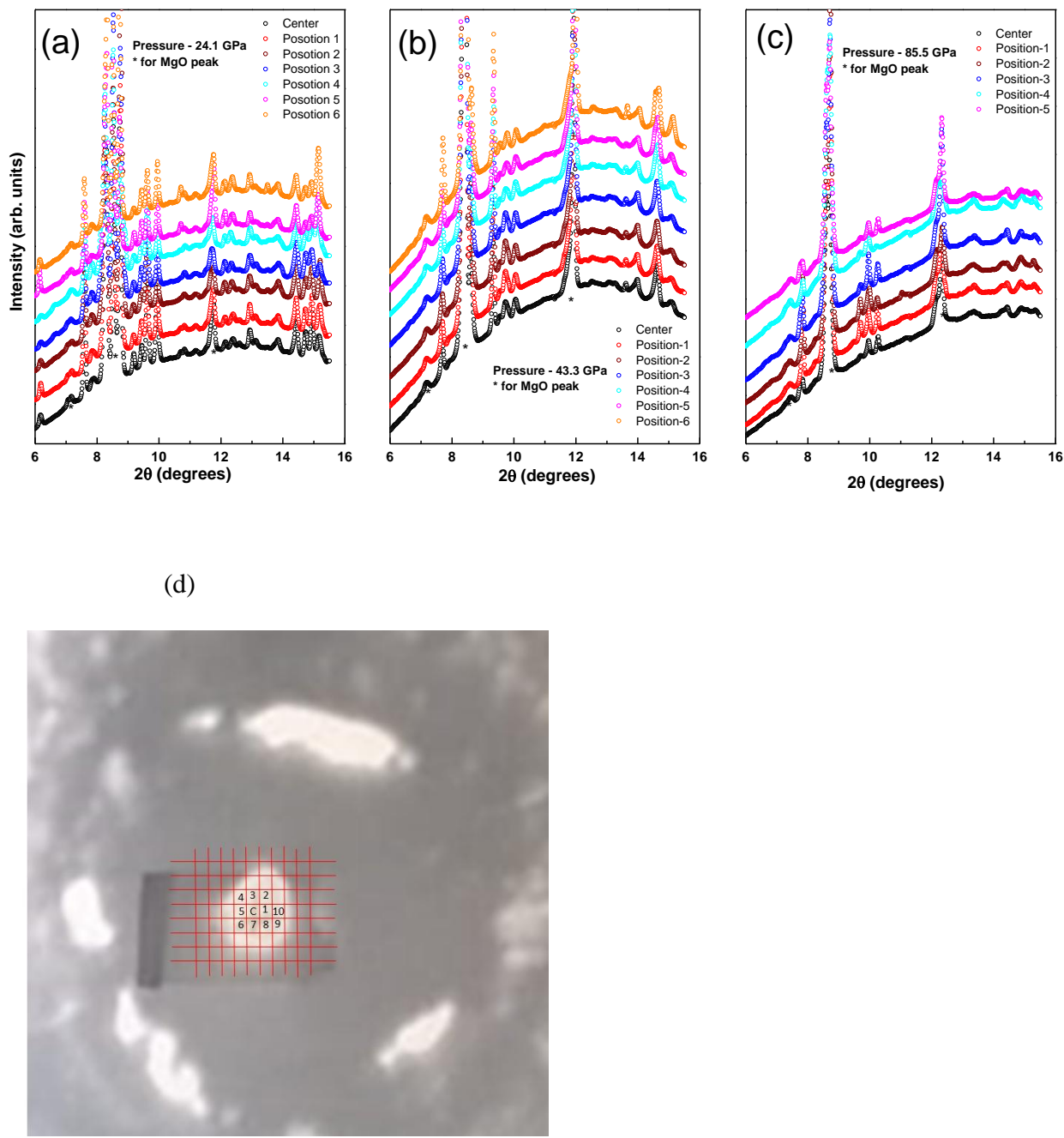

**Figure S11:** XRD patterns collected as grid run at three different pressure points: (a) and (b) for Run-2, and (c) for Run-4. A typical view of the positions from where XRD data are collected during grid run is shown in (d), C represent center position. This photograph was taken during heating in Run-2. Bright portion at the centre is the hotspot. We have collected xrd patterns from centerof the hot spot to outer regions during grid run.

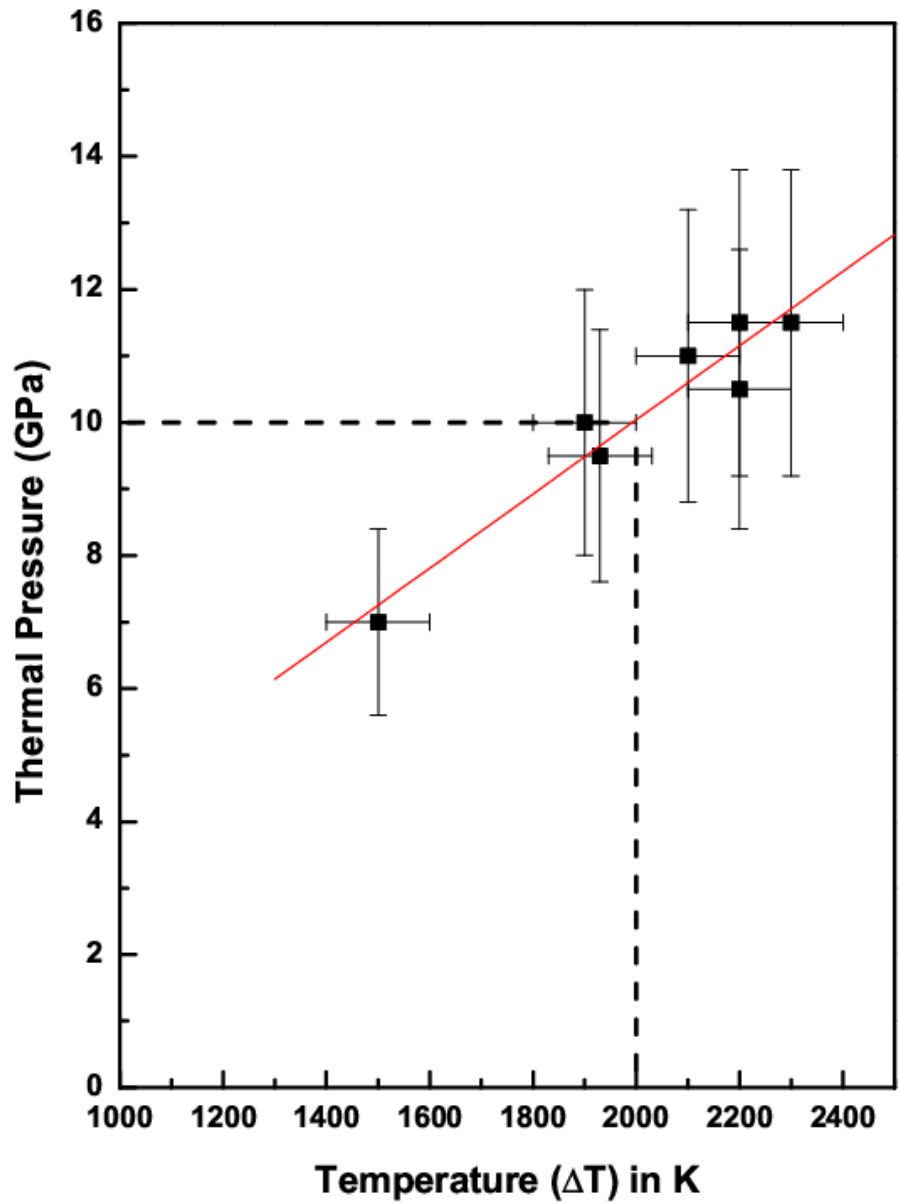

**Figure S12:** Thermal pressure ($P_{th}(T) = \alpha K_T \Delta T$) with temperature ($\Delta T = T - T_0$). Red solid fitted line yields $\alpha K_T = 0.0058\ (5)\ \text{GPa}^{-1}$.